\documentclass[preprint,showpacs,preprintnumbers,amsmath,amssymb,prb]{revtex4}

\usepackage{graphicx}
\usepackage{dcolumn}
\usepackage{bm}

\newdimen\tableauside\tableauside=1.0ex
\newdimen\tableaurule\tableaurule=0.4pt
\newdimen\tableaustep
\def\phantomhrule#1{\hbox{\vbox to0pt{\hrule height\tableaurule width#1\vss}}}
\def\phantomvrule#1{\vbox{\hbox to0pt{\vrule width\tableaurule height#1\hss}}}
\def\sqr{\vbox{%
  \phantomhrule\tableaustep
  \hbox{\phantomvrule\tableaustep\kern\tableaustep\phantomvrule\tableaustep}%
  \hbox{\vbox{\phantomhrule\tableauside}\kern-\tableaurule}}}
\def\squares#1{\hbox{\count0=#1\noindent\loop\sqr
  \advance\count0 by-1 \ifnum\count0>0\repeat}}
\def\tableau#1{\vcenter{\offinterlineskip
  \tableaustep=\tableauside\advance\tableaustep by-\tableaurule
  \kern\normallineskip\hbox
    {\kern\normallineskip\vbox
      {\gettableau#1 0 }%
     \kern\normallineskip\kern\tableaurule}%
  \kern\normallineskip\kern\tableaurule}}
\def\gettableau#1 {\ifnum#1=0\let\next=\null\else
  \squares{#1}\let\next=\gettableau\fi\next}

\newcommand{\xiv}{\mbox{\boldmath$\xi$}}
\newcommand{\sigmav}{\mbox{\boldmath$\sigma$}}
\newcommand{\tauv}{\mbox{\boldmath$\tau$}}

\begin{document}

\title{Inclusion of phonon exchange in a nuclear matrix element }

\author{Irfan Chaudhary}
\email{irfanc@mit.edu}
\author{Peter Hagelstein}
\email{plh@mit.edu}
\affiliation{Research Laboratory of Electronics \\ 
Massachusetts Institute of Technology \\
Cambridge, MA 02139,USA}
\pacs{63.10.+a, 63.20.-e, 63.20.Kr, 63.90.+t, 25.10.+s, 02.20.-a}

\begin{abstract}
Phonon exchange with nuclei in the course of fusion reactions
that occur in a solid have not been analyzed previously.  
This problem has become of interest in connection with claims of observations of
anomalies in metal deuterides.  
If the strong force interaction were dependent only on position (and not spin or isospin), then the coupling
with phonons can be developed directly.  
Since a nuclear interaction can change the lattice constituents, the initial and final state lattices
can be different, and we must include this in the formulation.  
For more realistic strong force models with spin and isospin dependence, we can use correlated nuclear wavefunctions which
are made up of products of space, spin and isospin components.  
In this case, the spin and isospin algebra can be done analytically, producing
channel-dependent potentials that are only space dependent.
The formulation that results can be used for quantitative estimates of phonon exchange.
\end{abstract}

\maketitle

\section{Introduction}

   Few claims have resulted in as much controversy as the claim of 
the observation of an excess heat effect of nuclear origin in metal deuterides 
made by Fleischmann, Pons, and Hawkins in 1989.\cite{FleischmannPons}  
Reviews of work on this problem were conducted by the Department of 
Energy (in the United States) in 1989\cite{ERAB} and in 2004.\cite{DoE2004}
The conclusions of the 1989 review were very negative.  
The 2004 review was quite different in how it was carried out, what material was
reviewed, and in the results.
On the specific question of the existence of an excess power effect, which was
the central issue of the review material presented to DoE,\cite{ReviewDoc}
there were more positive comments than negative comments volunteered from the reviewers.\cite{reviewcomments}
This is significant, as DoE did not charge the reviewers to consider this question
specifically.
Among many other comments that were made, 
several of the reviewers suggested that more work on this problem
be submitted to mainstream journals.

   Our interest in this paper is not focused on the question of the existence of
an excess heat effect, although this issue has motivated our investigations.  
We are interested instead in theoretical issues which must be better understood in
order for theory to be more relevant to the problem.  
For most physicists, the theoretical problem was adequately addressed by
Huizenga, who argued that three ``miracles'' were required for a theoretical
explanation.\cite{Fiasco}  
\begin{itemize}
\item The first ``miracle'' involves the question of reaction rate. 
The excess heat effect implies reaction rates on the order of $10^{12}$ sec$^{-1}$, 
tens of orders of magnitude larger than the fastest fusion rates possible in 
molecular D$_2$ or HD.  
\item The second ``miracle'' concerns the branching ratio.  Fleischmann initially
speculated that $^4$He was being made from two deuterons.  Such a result is
seemingly miraculous as the branching ratio for this reaction channel is lower
by about seven orders of magnitude than the primary n+$^3$He and p+t reaction
channels.
\item The third ``miracle'' involves the requirement that the reaction energy be
expressed through channels not involving energetic reaction products. 
\end{itemize}
If one believes that these issues are insurmountable, then one would require that
any claim of excess heat of nuclear origin must be accompanied by measurable energetic
reaction products.
This view was adopted by Huizenga, and by the 1989 ERAB Panel.  This view is also 
widely held among the physics community at this time.

\subsection{Motivation}

However, other views are possible.  
Fleischmann's initial speculation was that a new physical process was involved, one 
that behaves differently from what is described in the textbooks.  
It seems that if we are trying to understand whether an excess heat effect can exist,
we should be asking whether there can be interactions between nuclei and a condensed
matter environment.
For example, suppose that phonons are exchanged in the course of a nuclear reaction
that occurs in a lattice.
Could this phonon exchange result in a modification of the selection rules? 
%
%
And if phonon exchange occurs, would that not constitute a new channel into which
reaction energy might go?
%
%
%
   Nuclear physics relies heavily on the notion of an equivalent vacuum reaction
in order to understand quantitatively fusion reactions.  Huizenga's three ``miracles''
reflect such a view.  
For example, the second ``miracle'' is essentially a statement
of energy, linear momentum, and angular momentum conservation, 
all of which are conserved in vacuum.  
Is it obvious that conservation laws in a condensed matter environment
always produce the same reaction pathways with the same branching ratios;
and, if so, can this be proven?
In the case of the third ``miracle,'' the reaction energy must be expressed
as an energetic product in vacuum, since no other kinds of channels exist.
But is it obvious that this must also be the case in a condensed matter
environment, where low energy channels do exist?  Once again, can this be
proven?
In any event, all of these questions ought to be capable of being addressed by
theory.

The current lack of such a theory in the mainstream literature is underscored 
in comments made by some reviewers in the 2004 DoE review.
For example, a reviewer with expertise in metal deuterides and phonons wrote: 
``To create a coupling between nuclear interaction and phonons at such a 
low energy region (namely, the electromagnetic interaction) is beyond one's 
imagination at the moment.''
Another reviewer with expertise in nuclear theory wrote:
``I am convinced that simple order-of-magnitude estimates of this kind could 
quickly rule out any of the exotic mechanisms proposed...'' This reviewer
argues that the matrix element for the emission of a large number of
phonons would scale as $(kR)^{2L}$, where $k$ is the phonon wavenumber.
Aside from the fact that the emission of a large number of phonons in a
single process was not put forth as a proposed reaction mechanism in 
the review, the use (or misuse) by this reviewer of a phonon wavenumber in this 
way underscores the need for a relevant theory.  
We find ourselves then in a position where a better understanding of phonon
exchange in fusion reactions is required for a sensible discussion of these 
and related issues.  
This motivates us to consider in this manuscript how to calculate phonon exchange 
in perhaps the simplest application; that of an interaction matrix element.
Phonon exchange due to recoil in neutron capture reactions was 
analyzed by Lamb,\cite{Lamb} and this analysis has been used
subsequently for phonon exchange in gamma reactions.
In the M\"ossbauer effect, an anomalous zero-phonon exchange effect occurs if the
recoil is sufficiently gentle.\cite{Mossbauer}
In our matrix elements recoil is involved implicitly, and phonon
exchange due to the local change in lattice structure will
appear explicitly.

\subsection{Excitation transfer scheme}

   In the following sections of this manuscript, we develop a formulation suitable
for including phonon exchange in nuclear reaction matrix elements.
Once armed with this tool, we are in a position to develop calculations relevant
to candidate reaction schemes that involve phonon exchange (of which quite a few
have been proposed over the years).
Our current interest is focused on a particular scheme in which coupling occurs between
nuclei at different sites as a result of phonon exchange with a common phonon
mode.
In this scheme, two deuterons fuse locally with the reaction energy transferred elsewhere
through a second-order off-resonant excitation transfer effect.
The systematics of this effect in the idealized case of two-level systems coupled
to a common oscillator is discussed in a recent analysis.\cite{PLHandIUC}
In essence, the reaction energy is transferred to excited states in host lattice nuclei
through an initial slow excitation transfer step, and then subsequent rapid excitation
transfer occurs among these nuclei.
The reaction energy is transferred to the phonon mode in this scheme a few
phonons at a time in a very large number of fast excitation transfer reactions.
The scheme seems to be closely related to experiment, and the reaction rate
predicted appears to be consistent with experiment to within uncertainties 
in the deuteron-deuteron screening.
The evaluation of this scheme requires the ability to evaluate phonon exchange
in fusion reactions as discussed in the present manuscript.

Huizenga's arguments are based ultimately on fundamental notions relevant to
vacuum reactions, local energy and momentum conservation, and the use of
Golden Rule reaction rate estimates, as discussed above.
In this new excitation transfer scheme, the energy of the coupled system is
conserved, but local energy is transferred elsewhere.
In addition, the associated reaction dynamics do not follow simple Golden
Rule predictions.
Huizenga's arguments simply do not apply to this kind of scheme.

\subsection{Overview of the paper}

This paper is organized as follows.  In section II we discuss the basic
lattice-nuclear coupling (assuming a simplified nuclear potential model
that depends only on position),  
and argue that it is very similar to the more familiar electron-phonon and 
neutron-phonon couplings. 
However, realistic nuclear potentials depend on spin, isospin, and parity.
In Section III we show that for realistic nuclear potentials, by
first doing the spin and isospin
algebra analytically, the ideas developed in Section II are relevant. 
In Section IV, using the Hamada-Johnston nuclear potential, 
we explicitly carry out the program outlined in Section III for one particular 
matrix element relevant to a deuteron-deuteron fusion reaction.  This
particular interaction is chosen because it is one of the simplest realistic
nuclear potentials.  
In Section V we extend this vacuum result to include the lattice. 
We find in Section VI an explicit expression for one of the integrals
that appears in the interaction matrix element in terms of phonon coordinates.
This result is closely related to the discussion in Section II.
A summary and conclusions is given in Section VII.

Although the essential issues involved in phonon exchange can be seen
with the simplified position-dependent potential, we need to work with
nuclear wavefunctions that depend on spin and isospin when using a
more realistic nuclear potential model.  The construction of these
general four-body correlated nuclear wavefunction requires the use of
certain aspects of representation theory, and may 
be  unfamiliar to some of the readers.  As a result, we have included material
in a set of appendices that provide further discussion of the 
wavefunction construction and usage in matrix element calculations.


\newpage

\section{Phonon interactions}

   We are interested in the basic issue of phonon exchange in association
with nuclear reactions in a lattice.  The formulation that follows in
subsequent sections is complicated, and it makes sense to review phonon
exchange in more familiar circumstances.  
In condensed matter physics, phonon exchange is of interest in the study of 
electron scattering and neutron scattering.  In both cases, phonon exchange
comes about through a formulation in which the center of mass coordinates of
the nuclei are taken to be phonon operators.

\subsection{Phonon exchange associated with electronic transitions}

For example, electron-phonon interactions can be developed ultimately starting from
the Coulomb interaction written in the form\cite{Callaway}

\begin{equation}
-\sum_{j,k}
~{Z_j e^2 \over |\hat{\bf R}_j - {\bf x}_k|}
\end{equation}

\noindent
where the ${\bf x}_k$ are the electron coordinates, and the $Z_j$ are
the effective charges on the ions. The $\hat{\bf R}_j$
are the nuclear center of mass coordinates, which are dynamical, and
which we think of in terms of lattice operators ($\hat{q}_m$)
\begin{equation}
\hat{\bf R}_j ~=~ {\bf R}_j(\hat{q}_1, \hat{q}_2, \cdots )
\end{equation}

\noindent
In the event that we adopt a description of the lattice in terms of product
states, then a matrix element of the Coulomb interaction might be written
as

\begin{equation}
M_{fi}
~=~
-
\left \langle 
\Psi^L_f (\lbrace q \rbrace ) 
\psi_f (\lbrace {\bf x} \rbrace ) 
\left |
\sum_{j,k}
~{Z_j e^2 \over |\hat{\bf R}_j - {\bf x}_k|}
\right |
\Psi^L_i (\lbrace q \rbrace ) 
\psi_i (\lbrace {\bf x} \rbrace ) 
\right \rangle
\end{equation}

\noindent
In this way of thinking about the problem, we could define a phonon interaction 
by integrating over electronic coordinates

\begin{equation}
\hat{v}_{fi}(\lbrace q \rbrace ) 
~=~
-
\left \langle 
\psi_f (\lbrace {\bf x} \rbrace ) 
\left |
\sum_{j,k}
~{Z_j e^2 \over |\hat{\bf R}_j - {\bf x}_k|}
\right |
\psi_i (\lbrace {\bf x} \rbrace ) 
\right \rangle
\end{equation}

\noindent
Phonon exchange in this case can be developed then from lattice matrix elements of
this operator

\begin{equation}
M_{fi} 
~=~ 
\left \langle 
\Psi^L_f (\lbrace q \rbrace ) 
\left |
\hat{v}_{fi}(\lbrace q \rbrace ) 
\right |
\Psi^L_i (\lbrace q \rbrace ) 
\right \rangle
\end{equation}

\noindent
In the event that the electronic transition results in a change in the local force
constants, the initial and final state lattice may differ enough that this should
be reflected in the matrix element.

\subsection{Phonon exchange associated with neutron interactions}

A similar approach can be used in the case of neutron scattering.  In the case
of a simple position-dependent interaction

\begin{equation}
V_0 
\sum_{j}
~\delta^3 (\hat{\bf R}_j - {\bf r}_n)
\end{equation}

\noindent
we can develop the phonon interaction according to

\begin{equation}
\hat{v}_{fi}(\lbrace q \rbrace ) 
~=~
\left \langle 
\psi_f ({\bf r}_n) 
\left |
V_0 
\sum_{j}
~\delta^3 (\hat{\bf R}_j - {\bf r}_n)
\right |
\psi_i ( {\bf r}_n ) 
\right \rangle
\end{equation}

\noindent
In these equations ${\bf r}_n$ is the neutron coordinate.  With a knowledge of
the phonon interaction, we can analyze phonon exchange using $M_{fi}$
matrix elements as calculated above.

\subsection{Phonon exchange in a simplified model for nuclear reactions}

We might choose to work with an approximate position-dependent nuclear interaction model that
is a function of spatial coordinates alone

\begin{equation}
\sum_{\alpha < \beta}
v_n(|{\bf r}_\alpha - {\bf r}_\beta|)
\end{equation}

\noindent
where $\alpha$ and $\beta$ refer to nucleon (neutron and proton) coordinates.  The
interaction matrix element in this case can be written as

\begin{equation}
M_{fi}
~=~
\left \langle
\Psi^L_f( \lbrace {\bf r} \rbrace )
\left |
\sum_{\alpha < \beta}
v_n(|{\bf r}_\alpha - {\bf r}_\beta|)
\right |
\Psi^L_i( \lbrace {\bf r} \rbrace )
\right \rangle
\end{equation}

\noindent
where the integration here is taken over individual nucleon coordinates.  A nuclear reaction
may result in the initial lattice being different than the final lattice.  For example, if
two deuterons reacted to form $^3$He+n, the initial lattice would contain the two deuterons
while the final lattice would have $^3$He and a possibly free neutron.  We note that in
a normal version of this reaction, the final state products would fly off as energetic
nuclear products; we consider this possibility to be included within the formulation under
discussion.

However, this description does not yet make clear the situation with respect to phonon
exchange.  For this, we need to rewrite the integral in terms of phonon coordinates
rather than in terms of nucleon coordinates.  To make progress, we can express 
individual nucleon coordinates that are associated with nuclei in terms of the nuclear
center of mass coordinates ${\bf R}$ and relative internal nuclear coordinates $\xiv$

\begin{equation}
\Psi^L(\lbrace {\bf r} \rbrace )
~=~
\Psi^L(\lbrace {\bf R} \rbrace, \lbrace \xiv \rbrace )
\end{equation}

\noindent
Since a nuclear reaction may result in a change in nuclei between the initial states and the
final states, the set of center of mass coordinates may be different.  Consequently, the
nuclear interaction matrix element can be written formally as

{\small

\begin{equation}
M_{fi}
=
\int
\left[\Psi^L_f(\lbrace {\bf R}^f \rbrace, \lbrace \xiv^f \rbrace ) \right]^*
\sum_{\alpha < \beta}
v_n(|{\bf r}_\alpha - {\bf r}_\beta|)
\Psi^L_i(\lbrace {\bf R}^i \rbrace, \lbrace \xiv^i \rbrace )
\prod_\alpha \delta^3({\bf r}_\alpha^i-{\bf r}_\alpha^f)
d \lbrace \xiv^i \rbrace 
d \lbrace \xiv^f \rbrace 
d \lbrace {\bf R}^i \rbrace 
d \lbrace {\bf R}^f \rbrace 
\end{equation}
}


\noindent
We integrate over all initial and final state center of mass and internal coordinates, and
require that the individual nucleon coordinates be the same in the initial and final states.

Once we have a description in terms of the nuclear center of mass coordinates, we can
rewrite the integrations in terms of phonon coordinates.  
To simplify things, we might assume product states of the form

\begin{equation}
\Psi^L(\lbrace {\bf R} \rbrace, \lbrace \xiv \rbrace )
~\rightarrow~
\Psi^L(\lbrace q \rbrace) \Phi( \lbrace \xiv \rbrace )
\end{equation}

\noindent
In this case, the matrix element becomes

{\small
$$
M_{fi}
=
\ \ \ \ \ \ \ \ \ \ \ \ \ \ \ \ \ \ \ \ \ \ \ \ \ \ \ \ \ \ \ \ \ \ \ \ 
\ \ \ \ \ \ \ \ \ \ \ \ \ \ \ \ \ \ \ \ \ \ \ \ \ \ \ \ \ \ \ \ \ \ \ \ 
\ \ \ \ \ \ \ \ \ \ \ \ \ \ \ \ \ \ \ \ \ \ \ \ \ \ \ \ \ \ \ \ \ \ \ \ 
\ \ \ \ \ \ \ \ \ \ \ \ \ \ \ \ \ \ \ \ \ \ \ \ \ \ \ \ \ \ \ \ \ \ \ \ 
$$
\begin{equation}
\int
\left[ \Psi^L_f(\lbrace  q^f \rbrace) \Phi_f( \lbrace \xiv^f \rbrace )\right]^*
\sum_{\alpha < \beta}
v_n(|{\bf r}_\alpha - {\bf r}_\beta|)
\Psi^L_i(\lbrace q^i \rbrace) \Phi_i( \lbrace \xiv^i \rbrace )
\prod_\alpha \delta^3({\bf r}_\alpha^i-{\bf r}_\alpha^f)
d \lbrace \xiv^i \rbrace 
d \lbrace \xiv^f \rbrace 
d \lbrace q^i \rbrace 
d \lbrace q^f \rbrace 
\end{equation}

}


\noindent
For a final lattice with $N$ ions, this can be thought of as

\begin{equation}
M_{fi}
~=~
\int
\left[\Psi^L_f({\bf q}^f )\right]^* 
~\hat{v}_{fi}( {\bf q}^f , {\bf q}^i )~
\Psi^L_i( {\bf q}^i ) 
\delta^{(3N - 3)}({\bf q}^f - {\bf A}\cdot {\bf q}^i - {\bf b})
d {\bf q}^i  
d {\bf q}^f  
\end{equation}

\noindent
in which the phonon interaction operator $\hat{v}_{fi}({\bf q}^f
,{\bf q}^i)$ 
is developed from an appropriate integration over
the internal nuclear coordinates.  More simply, we may write
\begin{equation}
M_{fi}
~=~
\int
\left[\Psi^L_f({\bf q}^f )\right]^* 
~\hat{v}_{fi}( {\bf q}^f , {\bf q}^i )~
\Psi^L_i( {\bf q}^i ) 
d {\bf q}^i  
\label{FC}
\end{equation}

\noindent
together with the associated constraint 

\begin{equation}
{\bf q}^f ~=~ {\bf A} \cdot {\bf q}^i + {\bf b}
\end{equation}

\noindent
Due to the change in the lattice structure, the phonon mode structure can
be altered.\cite{Duschinsky, Sharp}  The associated matrix element written
in terms of initial state and final state coordinates [equation (\ref{FC})]
is common in papers on Franck-Condon factors in polyatomic molecules.\cite{Faulkner}

\subsection{Discussion}

   Phonon exchange in electron and neutron scattering can be implemented 
using the local nuclear center of mass operators as phonon operators. 
This leads to a straightforward description of matrix elements that include
phonon exchange, as well as the development of phonon exchange operators.
In the case of a nuclear reaction, we can define a nuclear interaction
matrix element simply enough under the assumption of a
position-dependent potential
as long as we work with nucleon coordinates.  However, when the integral
is rewritten in terms of phonon mode amplitudes, the resulting expression
is more complicated.  The underlying ideas and approach are very similar.
The important message here is that a nuclear interaction matrix element
expressed in terms of the nucleons of the lattice is completely straightforward,
and it is only the bookkeeping associated with the assembly of nucleons into
nuclei, and nuclei into phonons, that adds some complication to the problem.

There are basic differences between the new problem and the well known electron
and neutron scattering problems which deserve to be noted.  
\begin{itemize}
\item The energy scales in the case of nuclear reactions are not matched; 
    typical nuclear reaction energies are on MeV scale, 
    whereas the maximum phonon energies are typically 100 meV or less.  
\item A nuclear reaction involving a change in the nucleon content of
    the product nuclei will likely produce phonon exchange, if for no
    other reason than the fact that such reactions will produce a change
    in the structure of the lattice.  
\end{itemize}

\newpage

\section{Formal calculation with a realistic potential} 

For quantitative results, we will need to use a realistic nuclear
potential. 
In atomic physics and solid state physics, 
we are familiar with wavefunctions that have both spin and space dependence.
Nuclear wavefunctions have in addition isospin dependence.  
In the isospin scheme, a nucleon can be in an isospin down state
$|\downarrow \rangle$ in which case it is a neutron, or it can
be in an isospin up state $|\uparrow \rangle$ in which case it
is a proton.
It was recognized early on in nuclear physics that the nucleon-nucleon
potential is of short range in space, and depends explicitly on both 
spin and isospin.  
Over the past 60 years, increasingly accurate nuclear potential
models have been developed.  Early models were based in part on
field theoretical models for the one-pion exchange, and in part
on few parameter empirical models that were fit to scattering data.
In recent years, the most accurate nuclear potential models have
been constructed from a diagrammatic analysis of an effective field
theory.\cite{Machleidt,Eppelbaum}  
For the purposes of the present paper, we will adopt an
early potential model (the Hamada-Johnston potential\cite{HJ}), which 
has the advantage for our discussion of being relatively simple
in form.

\subsection{Hamada-Johnston potential}

The  Hamada-Johnston potential between nucleon 1 and nucleon 2 appears in the
literature written as 

\vskip -0.150in
\begin{equation}
V_{HJ}(1,2) 
~=~ 
V_C(1,2) ~+~ V_T(1,2) S_{12} ~+~ V_{LS}(1,2) ({\bf L}.{\bf S}) ~+~ V_{LL}(1,2) L_{12}
\end{equation}

\noindent 
This potential is made up of four basic terms: a central potential $V_C$;
a tensor interaction $V_T S_{12}$;
a spin-orbit term $V_{LS}({\bf L}.{\bf S})$; 
and a generalized centripetal potential term $V_{LL} L_{12}$.
These are discussed further in Appendix A.  The central and tensor
terms have isospin dependence, whereas the spin-orbit and generalized
centripetal terms do not.  As an example, we consider the tensor term in
more detail (which we use as an example in
Appendix~\ref{sec:specificmatelt}) . Explicitly it is given by 

\begin{equation}
V_T(1,2) S_{12} 
~=~ (\tauv_1 \cdot \tauv_2) \, y_T(\mu r_{12})  
\left[
 3\frac{(\sigmav_1 \cdot {\bf r}_{12}) \; (\sigmav_2 \cdot {\bf
     r}_{12})}{r_{12}^2} - \sigmav_1 \cdot \sigmav_2 
\right]
\end{equation}

\noindent
with $\mu = m_\pi c/\hbar$, where $m_\pi$ is the mass of the pion.
The radial potential is parameterized according to
\begin{equation}
y_T(x) 
~=~ 
0.08 {m_\pi c^2 \over 3} \left(1 + \frac{3}{x} + \frac{3}{x^2} \right) {e^{-x} \over x}
\left [
1 + 
a_T {e^{-x} \over x}
+ b_T {e^{-2x} \over x^2}
\right ]
\end{equation}

\noindent
where $a_T$ and $b_T$ are parameters which have 
been fit separately for singlet-triplet spin and even-odd parity
(angular momentum) channels.  Hence, $y_T$ has an additional implicit dependence
on spin and parity: $y_T = y_T^{\alpha}$ where $\alpha$ can be
even-triplet ($et$), even-singlet ($es$), odd-triplet($ot$) or
odd-singlet($os$).

\subsection{Wavefunctions}

Because realistic nuclear potentials have explicit dependences on
spin and isospin, one requires nuclear wavefunctions that have
well defined spin and isospin dependences.  In the nuclear physics
literature, there are a variety of approaches to this problem, 
ranging from determinantal wavefunctions built up from single
nucleon orbitals to more complicated correlated wavefunctions.
For the development of the results in this paper, we will
adopt a basic construction of correlated wavefunctions in
terms of products of spatial, spin-dependent, and isospin-dependent
pieces, each of which is determined by group theory.

As is customary in nuclear physics, we choose our wavefunctions to satisfy
the generalized Pauli exclusion principle,  
which requires that the total wavefunction (including space, spin and
isospin parts) has to be antisymmetric with respect to nucleon exchange.
A group theoretical construction of the antisymmetric wavefunctions is
possible by summing over products of
many-particle spatial functions, spin functions, and isospin functions,
each of which belong to appropriate representations of the symmetric 
group, $S(4)$.\cite{Mahmoud}
In the case of the spin and isospin functions, the symmetric group
representations correspond to eigenfunctions of total spin and isospin,
as a consequence of the Schur-Weyl duality.\cite{Elliott, Fulton}  
Hence the antisymmetric wavefunctions have well-defined total spin and
isospin. This is discussed in Appendix B.

In  general, such correlated wavefunctions can be written in the form

\begin{equation}
\Psi =  \sum_j C_j  s_j( \lbrace \sigma \rbrace ) t_j(\lbrace \tau \rbrace ) 
   \psi_j(\lbrace {\bf r} \rbrace ) 
\end{equation}

\noindent
where $s_j( \lbrace \sigma \rbrace )$ are the spin functions, 
$t_j(\lbrace \tau \rbrace )$ are the isospin functions, and
$\psi_j(\lbrace {\bf r} \rbrace )$ are the spatial functions; all
of which belong to certain representations of the symmetric group as
indicated above.
The $C_j$ are products of the Clebsch-Gordan
coefficients of $S(4)$. 

\subsection{Formal matrix element calculation}

Given the discussion above, the basic calculation of an interaction matrix element 
is straightforward.    
We begin with initial and final states defined as  

\vskip -0.150in
\begin{equation}
\Psi  
=   
\sum_j C_j  
s_j( \lbrace \sigma \rbrace ) 
t_j(\lbrace \tau \rbrace ) 
\psi_j(\lbrace {\bf r} \rbrace ) 
\ \ \ \ \ \ \ \ \ \ 
\Psi^\prime  
=  
\sum_k C_k  
s^\prime_k( \lbrace \sigma \rbrace ) 
t^\prime_k(\lbrace \tau \rbrace ) 
\psi^\prime_k(\lbrace {\bf r} \rbrace ) 
\end{equation}

\noindent 
We then formally calculate the matrix element to give 
\vskip -0.150in

{\small
\begin{samepage}

$$
\left \langle \Psi \left \vert \sum_{\alpha<\beta}V_{HJ}(\alpha,\beta) \right \vert \Psi^\prime \right \rangle
~=~
\ \ \ \ \ \ \ \ \ \ \ \ \ \ \ \ \ \ \ \ \ \ \ \ \ \ \ \ \ \
\ \ \ \ \ \ \ \ \ \ \ \ \ \ \ \ \ \ \ \ \ \ \ \ \ \ \ \ \ \
\ \ \ \ \ \ \ \ \ \ \ \ \ \ \ \ \ \ \ \ \ \ \ \ \ \ \ \ \ \
\ \ \ \ \ \ \ \ \ \ \ \ \ \ \ \ \ \ \ \ \ \ \ \ \ \ \ \ \ \
$$
$$
\left \langle 
\sum_j C_j  
s_j( \lbrace \sigma \rbrace ) 
t_j(\lbrace \tau \rbrace ) 
\psi_j(\lbrace {\bf r} \rbrace ) 
\bigg |
\sum_{\alpha<\beta}V_{HJ}(\alpha,\beta)
\bigg |
\sum_k C_k  
s^\prime_k( \lbrace \sigma \rbrace ) 
t^\prime_k(\lbrace \tau \rbrace ) 
\psi^\prime_k(\lbrace {\bf r} \rbrace ) 
\right \rangle
$$
\begin{equation}
~=~
\sum_{j,k}
\left \langle 
\psi_j(\lbrace {\bf r} \rbrace ) 
\bigg |
V_{R}^{j,k}(\lbrace {\bf r} \rbrace ) 
\bigg |
\psi^\prime_k(\lbrace {\bf r} \rbrace ) 
\right \rangle
\end{equation}

\end{samepage}

}

\noindent
In writing this, we have integrated formally over spin and isospin coordinates to
develop the interaction matrix element into simpler spatial matrix elements
involving individually position-dependent potentials $V_R^{j,k}$ given by

\begin{equation}
V_{R}^{j,k}(\lbrace {\bf r} \rbrace ) 
~=~ C_j^* C_k
\left \langle 
s_j( \lbrace \sigma \rbrace ) 
t_j(\lbrace \tau \rbrace ) 
\bigg |
\sum_{\alpha<\beta}V_{HJ}(\alpha,\beta)
\bigg |
s^\prime_k( \lbrace \sigma \rbrace ) 
t^\prime_k(\lbrace \tau \rbrace ) 
\right \rangle
\end{equation}

\noindent
In this way we can reduce the spin and isospin dependent 
nuclear matrix element into a set of simpler matrix elements,
each one of which involves a position-dependent potential.

\newpage

\section{Basic vacuum matrix element calculation}

  Before attempting a calculation of a specific interaction matrix element in the
lattice case, it is appropriate to first examine the simpler equivalent
calculation in the vacuum case.  For this specific calculation, we will select
a four-body problem in which there are two deuterons in the initial
state, and in which the final state contains a three nucleon body ($^3$He or t) 
and a one nucleon body (n or p).

\subsection{Vacuum nuclear wavefunctions}

The specific matrix element that we select for our example involves
a quintet ($S = 2, M_S = 1$) initial state $\Psi_i$ with two deuterons 
(the deuterons each have spin 1), and a singlet ($S = 0,M_S = 0$)
final state which is a linear combination of $^3$H + p and $^3$He + n.
The reason for this is that a linear combination of states with different
total isospin ($T=0$ and $T=1$) are required to resolve a single proton 
or neutron in the final state channel.  For simplicity, we have adopted 
initial and final state wavefunctions with total isospin $T=0$.
As we discussed briefly above, nuclear wavefunctions can be constructed from
linear combinations of products of basis vectors of representations of
the symmetric group. 
For the case of four-body wavefunctions, we have indexed the specific basis vectors
in terms of their associated Yamanouchi symbol in Appendix B.  We have
also listed the results of a systematic construction of all such
four-body wavefunctions with total isospin zero in Appendix C.

The initial state is constructed according to

\begin{equation}
\Psi_i 
~=~ 
\Psi_6  
~=~  
\frac{1}{\sqrt{2}} 
\left[\psi_{5}  s_{10}t_6 -  \psi_{6}s_{10} t_5 \right]
\end{equation} 

\noindent
where $\Psi_6$ is the notation used in systematic construction of Appendix B.
Each of the four-particle spin ($s$), isospin ($t$), and spatial ($\psi$)
functions are individually basis vectors of representations of the symmetric 
group listed in Appendix B. 
The final state is constructed according to

\begin{equation}
\Psi_f 
~=~
\Psi_1 
~ = ~  
\psi_{10} 
\frac{1}{\sqrt{2}} 
\left[  s_5 t_6 - s_6  t_5 \right] 
\end{equation}

\noindent
where $\Psi_1$ is also the notation used in Appendix B.
We have selected these states in particular due to the close correspondence 
between the physical states discussed below, 
and the group theoretical Yamanouchi basis states for these channels.
In the case of other channels, a linear superposition of Yamanouchi
basis states may be required.

\subsection{Matrix element in terms of Yamanouchi basis functions}

The matrix element can be evaluated using the approach discussed above.
We first express the matrix elements in terms of $V_{HJ}(1,2)$, since the
Yamanouchi basis functions are either symmetric or antisymmetric under
$1 \leftrightarrow 2$.  We write

\begin{equation}
\left \langle \Psi_f \left 
\vert \sum_{\alpha<\beta}V_{HJ}(\alpha,\beta) 
\right \vert \Psi_i \right \rangle
~=~
6
\left \langle \Psi_f \left \vert V_{HJ}(1,2) \right \vert \Psi_i \right \rangle
\end{equation}

\noindent
Using the initial and final states in the form listed above, we obtain
four different contributions to the matrix element

\begin{samepage}

$$
M_{fi}
~=~
3 
\bigg [ 
\langle s_5 t_6 \psi_{10} | V_{HJ}(1,2) | s_{10}t_6 \psi_5 \rangle
-
\langle s_5 t_6 \psi_{10} | V_{HJ}(1,2) | s_{10}t_5 \psi_6 \rangle
\ \ \ \ \ \ \ \ \ \ \ \ \ \ \ \ \ \ \ \ \ \ \ \ \ \ \ \ \ \ \ \ 
\ \ \ \ \ \ \ \ \ \ \ \ \ \ \ \ \ \ \ \ \ \ \ \ \ \ \ \ \ \ \ \ 
$$
\vskip -0.250in
\begin{equation}
\ \ \ \ \ \ \ \ \ \ \ \ \ \ \ \ \ \ \ \ \ \ \ \ \ \ \ \ \
-
\langle s_6 t_5 \psi_{10} | V_{HJ}(1,2) | s_{10}t_6 \psi_5 \rangle
+
\langle s_6 t_5 \psi_{10} | V_{HJ}(1,2) | s_{10}t_5 \psi_6 \rangle
\bigg ]
\end{equation}

\end{samepage}

\noindent
Because of the symmetry of the interaction and the wavefunctions, only
one term survives the spin and isospin algebra. The details of this
calculation for the tensor term are worked out in  Appendix
~\ref{sec:specificmatelt}. The generalized centripetal term can be
worked out in a completely analogous way.  The net result is that 
\begin{equation}
M_{fi}
~=~
\langle \psi_{10} | V_{R}^{10,5}(1,2)|\psi_5 \rangle
\end{equation}

\noindent
where 


\begin{equation}
V_{R}^{10,5}(1,2)
~=~
18 \sqrt{3} \, \frac{(x_{12} + i y_{12})\, z_{12} \, }{r_{12}^2}y_T^{et}(\mu r_{12}) 
~+~
\frac{\sqrt{3}}{\hbar^2} (\hat{L}_{+} \hat{L}_{z} + \hat{L}_{z} \hat{L}_{+}) y_{LL}^{et}(\mu r_{12})
\end{equation}

\noindent
where $\hat{L}_z$ and $\hat{L}_+$ are relative 12 (read as one-two) angular momentum operators.

We have already noted that the Yamanouchi basis is naturally purely symmetric
or antisymmetric 
under the exchange $1 \leftrightarrow 2$ (This is further discussed in
Appendix~\ref{sec:specificmatelt}).  As a consequence, the
calculation of the many-body matrix element leads to the simplest algebraic
result when specified in terms of 12 terms.  Interactions between all particles
are represented, since the $\psi_{10}$ and $\psi_{5}$ representations include
permutations of the particle numbering.  Hence, the use of a
position-dependent interaction
potential given in terms of ${\bf r}_{12}$ is appropriate, and implies the
use of ${\bf r}_{12}$ as a favored relative coordinate to use in the evaluation
of the spatial matrix element.

\subsection{Physical wavefunctions}

  The group theoretical machinery that we used in the evaluation above
results in a compact expression given in terms of spatial Yamanouchi basis
functions.  As such, they
have certain symmetries with respect to particle interchange.  These must be
implemented in the construction of the spatial wavefunctions.  This can
be accomplished by using an appropriate superposition of permutations of
wavefunctions with no particular symmetry.  

For example, suppose we begin with an unsymmetrized ``physical'' wavefunction
for the two deuteron initial state 

\begin{equation}
\label{eq:psi12psi34}
\psi(12;34) ~=~ 
\phi_d({\bf r}_2 - {\bf r}_1)\, 
\phi_d({\bf r}_4 - {\bf r}_3) \,
F_{2,2} \left ( \frac{{\bf r}_1 + {\bf r}_2}{2}, 
          \frac{{\bf r}_3 + {\bf r}_4}{2} \right ) 
\end{equation}

\noindent
In this wavefunction, nucleons 1 and 2 make up one deuteron  
given by $\phi_d({\bf r}_2 - {\bf r}_1)$, 
and nucleons 3 and 4 make up another deuteron $\phi_d({\bf r}_4 - {\bf r}_3)$.
The generalized channel separation function 
$F_{2,2}$ is the probability amplitude associated with the two
center of mass coordinates of the deuterons.
As discussed in Appendix E, we can use this
unsymmetrized physical wavefunction as the basis 
for the construction of the associated Yamanouchi $\psi_5$ function,
which leads to

\begin{equation}
\psi_5 
~=~ 
\frac{1}{\sqrt{12}} 
\bigg [ 2 \psi(12;34) + 2 \psi(34;21) -
  \psi(23;14)   
  - \psi(14;23)  - \psi(13;24) -
  \psi(24;13) \bigg ] 
\label{eq:psi5}
\end{equation}

\noindent
This generates the appropriate Yamanouchi basis vector 
for use in the matrix element expression above.

  In the case of the final state channel, we begin with the physical wavefunction

\begin{equation}
\label{eq:psi123psi4}
\psi(123;4) ~=~
\phi_{3He}({\bf r}_3-{\bf r}_2,{\bf r}_2 - {\bf r}_1) \; 
F_{3,1} \left (\frac{{\bf r}_1 + {\bf r}_2 + {\bf r}_3}{3},{\bf r}_4 \right ) 
\end{equation}

\noindent
where $\phi_{3He}$ is the $^3$He wavefunction specified in terms of two relative internal
coordinates, and where $F_{3,1}$ is the generalized channel function for the center of
mass for the $^3$He and neutron.  From this physical wavefunction, we can construct
the associated Yamanouchi $\psi_{10}$ function according to

\begin{equation}
\label{eq:psi10}
\psi_{10}  
~=~ 
 \frac{1}{2} 
\bigg [
\psi(123;4) + \psi(124;3) +
  \psi(134;2) + \psi(234;1) 
\bigg ]
\end{equation} 

\noindent
This function can be used in the group theoretical matrix element formula above.

\newpage

\section{Lattice result}

Let us now consider exactly the same physical process taking place
inside the lattice.  We are interested in the calculation of the
interaction matrix element with wavefunctions that contain both
nuclear and lattice coordinates.  Perhaps the simplest way to
approach the problem is to make use of the vacuum results from
the last section, but expand the definition of the initial and
final states to include the lattice.

\subsection{Wavefunctions for lattice and nuclei}

  We now augment the definition of the physical wavefunctions to include
the lattice coordinates.  For example, in the new physical initial state wavefunction,
we include the center of mass coordinates of the other nuclei to obtain

\begin{equation}
\label{eq:Latpsi12psi34}
\psi(12;34,\{{\bf R}\}) ~=~ 
\phi_d({\bf r}_2 - {\bf r}_1)\, 
\phi_d({\bf r}_4 - {\bf r}_3) \,
F_{2,2}^L \left ( \frac{{\bf r}_1 + {\bf r}_2}{2}, 
          \frac{{\bf r}_3 + {\bf r}_4}{2},\{{\bf R}\} \right ) 
\end{equation}

\noindent
The other nuclei in the lattice are spectators in the sense that they do not 
participate in nucleon rearrangement associated with the reaction.  Consequently,
the appearance of the associated degrees of freedom into the lattice channel
separation factor is an appropriate generalization of the generalized channel
separation factor from the vacuum case.  We still require the nucleons that
participate in the reaction to be described by a wavefunction that is totally
antisymmetric under the exchange of nucleons.  The vacuum formulation that we
discussed above accomplished this in the absence of spectator nuclei.  There
seems to be no reason that we cannot adopt precisely the same kind of formulation
here (this is discussed in Appendix F).  
Hence, we form the appropriate Yamanouchi spatial basis function
for the initial state made up of the same superposition of physical states
that are augmented with spectator coordinates.  This leads to

$$
\psi_5^L 
~=~  
\frac{1}{\sqrt{12}} 
\bigg [ 
2 \psi(12;34,\{{\bf R}\}) +
  2 \psi(34;21,\{{\bf R}\})  -
  \psi(23;14,\{{\bf R}\})  
\ \ \ \ \ \ \ \ \ \ \ \ \ \ \ \ \ \ \ \ \ \ \ \ \ \ \ 
\ \ \ \ \ \ \ \ \ \ \ \ \ \ \ \ \ \ \ \ \ \ \ \ \ \ \ 
\ \ \ \ \ \ \ \ \ \ \ \ \ \ \ \ \ \ \ \ \ \ \ \ \ \ \ 
$$
\begin{equation}
\ \ \ \ \ \ \ \ \ \ \ \ \ \ \ \ \ \ \ \ \ \
- \psi(14;23,\{{\bf R}\})  - \psi(13;24,\{{\bf R}\}) -
  \psi(24;13,\{{\bf R}\}) \bigg ] 
\label{eq:Latpsi5}
\end{equation}

\noindent
Final state wavefunctions can be developed similarly.  The physical wavefunction
for the final state is

\begin{equation}
\psi(123;4,\{{\bf R}\}) 
~=~
\phi_{3He}({\bf r}_3-{\bf r}_2,{\bf r}_2 - {\bf r}_1) \; 
F_{3,1}^L \left (\frac{{\bf r}_1 + {\bf r}_2 +
  {\bf r}_3}{3},{\bf r}_4,\{{\bf R}\} 
\right )  
\label{eq:Latpsi123psi4}
\end{equation} 

\noindent
The appropriate final state Yamanouchi basis is formed through

\begin{equation}
\psi_{10}^L  
~=~  
\frac{1}{2} 
\bigg [ 
\psi(123;4,\{{\bf R}\}) 
+
\psi(124;3,\{{\bf R}\}) 
+ 
\psi(134;2,\{{\bf R}\}) 
+ 
\psi(234;1,\{{\bf R}\}) 
\bigg ]
\label{eq:Latpsi10}
\end{equation}

\subsection{Matrix element}

  We can take advantage of the same group theoretical construction to evaluate the
matrix element when spectator nuclei are present, since we only required total antisymmetrization
on the four nucleons involved in the local strong force interaction.  Consequently, we may write

\begin{equation}
M_{fi}
~=~
\langle \psi_{10}^L | V_{R}^{10,5}|\psi_5^L \rangle
\end{equation}

\noindent
We see in this result a very strong connection between the vacuum result and the lattice result.
In general we can systematically extend vacuum calculations to the lattice case through the
addition of the lattice nuclear center of mass coordinates as spectator degrees of freedom.

\subsection{Matrix element in terms of physical wavefunctions}

  We have made use of representations of the symmetric group in order to 
construct nuclear wavefunctions and to evaluate the interaction matrix
element.  In the case of the spatial wavefunctions, these basis functions
are made up of several different permutations of the physical wavefunction.
If we make use of these expansions, we can recast the interaction matrix
element in terms of matrix elements involving physical wavefunctions.  This
will in general produce a large number of terms.  

In the specific case of the tensor matrix element that we have been
using as an example, we obtain twenty four terms in total. We may write

\begin{samepage}

{\small

$$
M_{fi}
~=~
\frac{1}{\sqrt{12}} \int  
\left[ \phi_{3He}({\bf r}_3-{\bf r}_2,{\bf r}_2 - {\bf r}_1) 
\; 
F_{3,1}^L \left (\frac{{\bf r}_1 + {\bf r}_2 +
  {\bf r}_3}{3},{\bf r}_4,\{{\bf R}\}  \right ) \right]^* 
V_{R}^{10,5}({\bf r}_2 - {\bf r}_1)  
\phi_d({\bf r}_2 - {\bf r}_1)\, 
\phi_d({\bf r}_4 - {\bf r}_3)  
$$
\begin{equation}
 F_{2,2}^L \left ( \frac{{\bf r}_1 + {\bf r}_2}{2}, 
          \frac{{\bf r}_3 + {\bf r}_4}{2},\{{\bf R}\}  \right ) 
\; 
d^3 {\bf r}_1 \; 
d^3 {\bf r}_2 \; 
d^3 {\bf r}_3 \;
d^3 {\bf r}_4 
d^{3N-3}\{{\bf R}\} 
~+~ \cdots
\end{equation}

}

\end{samepage}

\noindent 
where the $\cdots$ represent the other twenty three terms in the
expansion (see Appendix G).

\newpage
\section{Matrix element in terms of phonon coordinates}

   A primary goal of this paper is to develop an expression for one of 
the integrals that appear in the matrix element explicitly in terms of
phonon coordinates, similar to what we discussed in Section II.  We begin
by defining the integral $I$ according to

{\small

$$
I
~=~
\int  
\left[ \phi_{3He}({\bf r}_3-{\bf r}_2,{\bf r}_2 - {\bf r}_1) 
\; 
F_{3,1}^L \left (\frac{{\bf r}_1 + {\bf r}_2 +
  {\bf r}_3}{3},{\bf r}_4,\{{\bf R}\}  \right ) \right]^* 
V_{R}^{10,5}({\bf r}_2 - {\bf r}_1)  
\phi_d({\bf r}_2 - {\bf r}_1)\, 
\phi_d({\bf r}_4 - {\bf r}_3)  
$$
\begin{equation}
 F_{2,2}^L \left ( \frac{{\bf r}_1 + {\bf r}_2}{2}, 
          \frac{{\bf r}_3 + {\bf r}_4}{2},\{{\bf R}\}  \right ) 
\; 
d^3 {\bf r}_1 \; 
d^3 {\bf r}_2 \; 
d^3 {\bf r}_3 \;
d^3 {\bf r}_4 
d^{3N-3}\{{\bf R}\} 
\end{equation}

}

\subsection{Integral in terms of center of mass and relative coordinates}

This integral includes integrations over nucleon coordinates as well as the 
center of mass coordinates of the spectator nuclei.  We would like to recast
the integral in terms of center of mass coordinates and relative coordinates,
to set things up for a transformation to phonon mode coordinates.
We may write


{\small

$$
I 
~=~ 
\int  
\left[ \phi_{3He}(\xiv^f_{32},\xiv^f_{21}) \; 
F_{3,1}^L\left ({\bf R}^f_{3He},{\bf r}^f_4,\{{\bf R}\} \right ) \right]^*
V(\xiv^f_{21}) 
\phi_d(\xiv^i_{21}) \, 
\phi_d(\xiv^i_{43}) \,
F_{2,2}^L \left({\bf R}^i_{d1}, {\bf R}^i_{d2},\{{\bf R}\} \right)
\ \ \ \ \ \ \ \ \ \ \ \ \ \
$$
$$
\delta^{(3)} \left( - \frac{1}{3} \xiv^f_{32} - \frac{2}{3} \xiv^f_{21}
+ {\bf R}^f_{3He} - ( - \frac{1}{2} \xiv^i_{21} + {\bf R}^i_{d1})
\right ) 
\delta^{(3)}\left( - \frac{1}{3} \xiv^f_{32} + \frac{1}{3}
\xiv^f_{21} + {\bf R}^f_{3He} - (\frac{1}{2} \xiv^i_{21} + {\bf
  R}^i_{d1}) \right) 
$$
$$
\delta^{(3)}\left( \frac{2}{3} \xiv^f_{32} + \frac{1}{3} \xiv^f_{21} +
      {\bf R}^f_{3He} - (-\frac{1}{2} \xiv^i_{43} + {\bf R}^i_{d2} )
      \right) 
\delta^{(3)}\left({\bf r}^f_4 - (\frac{1}{2} \xiv^i_{43} + {\bf
    R}^i_{d2}) \right) \; 
$$
\begin{equation}
d^3\xiv^f_{32} \; 
d^3\xiv^f_{21} 
d^3{\bf R}^f_{3He}\; 
d^3{\bf r}^f_4 \; 
~
d^3 \xiv^i_{21}\; 
d^3 \xiv^i_{43} \; 
d^3{\bf R}^i_{d1} \; 
d^3{\bf R}^i_{d2}  
~
d^{3N-3} \{{\bf R}\}
\end{equation}

}

\noindent
where we have defined the relative coordinates

\begin{equation}
\xiv^i_{21} = {\bf r}^i_2 - {\bf r}^i_1
\ \ \ \ \
\xiv^i_{43} = {\bf r}^i_4 - {\bf r}^i_3
\ \ \ \ \
\xiv^f_{32} = {\bf r}^f_3 - {\bf r}^f_2
\ \ \ \ \ 
\xiv^f_{21} = {\bf r}^f_2 - {\bf r}^f_1
\end{equation}

\noindent
and the center of mass coordinates

\begin{equation}
{\bf R}^i_{d1} = \frac{{\bf r}^i_1 + {\bf r}^i_2}{2}
\ \ \ \ \
{\bf R}^i_{d2} = \frac{{\bf r}^i_3 + {\bf r}^i_4}{2} 
\ \ \ \ \
{\bf R}^f_{3He} = \frac{{\bf r}^f_1 + {\bf r}^f_2 + {\bf r}^f_3}{3} 
\end{equation}

\subsection{Deuteron-deuteron separation}

  When the two deuterons are close enough for strong force interactions to
take place, then the Coulomb repulsion is sufficiently strong that this
interaction dominates over the forces from other atoms.  
Although we can use a phonon description in this case, it would not be the most natural,
and the Coulomb forces would require the presence of very high-order phonon
operators.  
Hence, it may be more convenient to adopt the point of view 
that the two deuterons appear to the lattice as an equivalent $^4$He
nucleus when they are a few fermis apart.  
The separation between them may best be handled then as part of the microscopic nuclear problem.
In this case, we may write

{\small
$$
I 
~=~
\int  
\left[ \phi_{3He}(\xiv^f_{32},\xiv^f_{21}) \; 
F_{3,1}^L\left ({\bf R}^f_{3He},{\bf r}^f_4,\{{\bf R}\} \right ) \right]^*
V(\xiv^f_{21}) 
\phi_d(\xiv^i_{21})\, 
\phi_d(\xiv^i_{43})\,  
\ \ \ \ \ \ \ \ \ \ \ \ \ \ \ \ \ \ \ \ \ \ \ \ \ \ \ \ \ \ \ \ \ \ \ \ \
$$
$$
F_{2,2}^L \left( {\bf R}^i_{4He} - \frac{1}{2} {\bf r}^i_{dd}, {\bf
  R}^i_{4He} + \frac{1}{2} {\bf r}^i_{dd}, \{{\bf R}\} \right)
\delta^{(3)} \left( - \frac{1}{3} \xiv^f_{32} - \frac{2}{3} \xiv^f_{21}
+ {\bf R}^f_{3He} - ( - \frac{1}{2} \xiv^i_{21} + {\bf R}^i_{4He} -
\frac{1}{2} {\bf r}^i_{dd}) 
\right) 
$$
$$
\delta^{(3)}\left( - \frac{1}{3} \xiv^f_{32} + \frac{1}{3}
\xiv^f_{21} + {\bf R}^f_{3He} - (\frac{1}{2} \xiv^i_{21} + {\bf
  R}^i_{4He} - \frac{1}{2} {\bf r}^i_{dd}) \right)   
$$
$$
\delta^{(3)}\left( \frac{2}{3} \xiv^f_{32} + \frac{1}{3} \xiv^f_{21} +
      {\bf R}^f_{3He} - (-\frac{1}{2} \xiv^i_{43} + {\bf R}^i_{4He} +
      \frac{1}{2} {\bf r}^i_{dd} ) 
      \right) 
$$
\begin{equation}
\delta^{(3)}\left({\bf r}^f_4 - (\frac{1}{2} \xiv^i_{43} + {\bf
    R}^i_{4He} + \frac{1}{2} {\bf r}^i_{dd}) \right) \;
d^3\xiv^f_{32} \; 
d^3\xiv^f_{21} 
d^3{\bf R}^f_{3He}\; 
d^3{\bf r}^f_4 \; 
d^3 \xiv^i_{21} \; 
d^3 \xiv^i_{43} \; 
d^3{\bf R}^i_{4He} \; 
d^3 {\bf r}^i_{dd}  
d^{3N-3}\{{\bf R}\} 
\label{eq:LatTransformedMatElt2}
\end{equation}

}

\noindent
In writing this, we have made use of the initial state relative and center of mass coordinates

\begin{equation}
{\bf R}^i_{4He} = \frac{1}{2} \left( {\bf R}^i_{d1} + {\bf R}^i_{d2} \right) 
\ \ \ \ \ 
{\bf r}^i_{dd} = {\bf R}^i_{d2} - {\bf R}^i_{d1} 
\end{equation}

  Note that in casting the matrix element in this way we are not assuming that the microscopic $^4$He
nuclear wavefunction is the same as the microscopic nuclear wavefunction for two deuterons.  
Instead, we recognize in these expressions that there is little difference from the point of view
of the rest of the lattice between two deuterons localized on the fermi scale (which the rest of the lattice sees as
a charge 2 and mass 4 object), and a $^4$He nucleus (which the rest of the lattice also sees as
a charge 2 and mass 4 object).
Due to the Coulomb repulsion between the two deuterons when they are close, the associated relative
wavefunction is far into a tunneling regime.
We have the choice to describe this tunneling in terms of deuteron coordinates as phonon operators, 
which is complicated, or else to describe this tunneling via a relative nuclear coordinate based on
a $^4$He center of mass coordinate, which is much simpler.
The formal problem is not changed with this replacement, although the approximations used in the
evaluation of the resulting expressions could be different.

\vskip 0.5in

\subsection{Transformation to phonon coordinates}

We are now in a position to make the transformation to phonon coordinates.
In the initial state, a nuclear center of mass coordinates can be expressed
in terms of phonon coordinates in general according to

\begin{equation}
\hat{\bf R}_j^i ~=~ {\bf R}_j^{i,0} ~+~ \sum_m {\bf u}^i_j[m] \hat{q}^i_m
\end{equation}

\noindent
Consequently, we may transform from initial state center of mass coordinates
to initial state phonon coordinates

\begin{equation}
{\bf R}_{4He}^i, \{{\bf R}\}
~~\rightarrow~~
{\bf R}_L^i, \{q^i\}
\end{equation}

\noindent
where ${\bf R}_L^i$ is the center of mass of the initial state lattice.  A similar
transformation is possible for the final state center of mass coordinates, 
which we indicate by

\begin{equation}
{\bf R}_{3He}^f, \{{\bf R}\}
~~\rightarrow~~
{\bf R}_L^f, \{q^f\}
\end{equation}

\noindent
where ${\bf R}_L^f$ is the center of mass of the final state lattice.

$$
I ~=~
\int  
\left[ \phi_{3He}(\xiv^f_{32},\xiv^f_{21}) \; 
F_{3,1}^L\left (\hat{{\bf R}}^f_{3He},{\bf r}^f_4,\{\hat{{\bf R}}^f\}
\right ) \right]^* 
V(\xiv^f_{21}) 
\phi_d(\xiv^i_{21})\, 
\phi_d(\xiv^i_{43}) 
\ \ \ \ \ \ \ \ \ \ \ \ \ \ \ \ \ \ \ \ \ \ \ \ \ \ \ \ \ \ \ \ \ \ \
$$
$$
F_{2,2}^L \left( \hat{{\bf R}}^i_{4He} - \frac{1}{2} {\bf r}^i_{dd}, \hat{{\bf
  R}}^i_{4He} + \frac{1}{2} {\bf r}^i_{dd}, \{\hat{{\bf R}}^i\} \right)
\delta^{(3)}\left({\bf R}^f_L - {\bf R}^i_L \right)
$$
$$
\delta^{(3)} \left( - \frac{1}{3} \xiv^f_{32} - \frac{2}{3} \xiv^f_{21}
+ \hat{{\bf R}}^f_{3He} - ( - \frac{1}{2} \xiv^i_{21} + \hat{{\bf R}}^i_{4He} -
\frac{1}{2} {\bf r}^i_{dd}) 
\right) 
$$
$$
\delta^{(3)}\left( - \frac{1}{3} \xiv^f_{32} + \frac{1}{3}
\xiv^f_{21} + \hat{{\bf R}}^f_{3He} - (\frac{1}{2} \xiv^i_{21} + \hat{{\bf
  R}}^i_{4He} - \frac{1}{2} {\bf r}^i_{dd}) \right) 
$$
$$  
\delta^{(3)}\left( \frac{2}{3} \xiv^f_{32} + \frac{1}{3} \xiv^f_{21} +
      \hat{{\bf R}}^f_{3He} - (-\frac{1}{2} \xiv^i_{43} + \hat{{\bf
      R}}^i_{4He} + 
      \frac{1}{2} {\bf r}^i_{dd} ) 
      \right) 
$$
$$
\delta^{(3)}\left({\bf r}^f_4 - (\frac{1}{2} \xiv^i_{43} + \hat{{\bf
    R}}^i_{4He} + \frac{1}{2} {\bf r}^i_{dd}) \right) \;
      \delta^{(3N-3)}( {\bf q}^f - \bar{\bf A} \cdot {\bf q}^i - {\bf b} ) 
$$
\begin{equation}
d^3{\bf R}^f_L \;
d^3{\bf R}^i_L  
d^3\xiv^f_{32} \; 
d^3\xiv^f_{21} \; 
d^3{\bf r}^f_4 \; 
d^3 \xiv^i_{21}\; 
d^3 \xiv^i_{43} \; 
d^3 {\bf r}^i_{dd}  
d^{3N-3} {\bf q}^f
d^{3N-3} {\bf  q}^i
\end{equation}

\noindent 
where

\begin{eqnarray}
\hat{{\bf R}}^f_{3He} & = & \hat{{\bf R}}^f_{3He}\left({\bf
  q}^f,{\bf R}^f_L \right) \\
\hat{{\bf R}}^f_k & = & \hat{{\bf R}}^f_k \left({\bf q}^f,
  {\bf R}^f_L \right) \\
\hat{{\bf R}}^i_{4He} & = &\hat{{\bf R}}^i_{4He} \left({\bf q}^i,
  {\bf R}^i_L \right) \\ 
\hat{{\bf R}}^i_k & = & \hat{{\bf R}}^i_k \left({\bf
  q}^i,{\bf R}^i_L \right) 
\end{eqnarray}

\subsection{Integral in terms of generalized lattice wavefunctions}

The generalized lattice channel separation factors are the same as lattice wavefunctions
augmented with additional degrees of freedom.  Let us adopt a notation of the form

\begin{equation}
F_{2,2}^L \left( \hat{{\bf R}}^i_{4He} - \frac{1}{2} {\bf r}^i_{dd}, \hat{{\bf
  R}}^i_{4He} + \frac{1}{2} {\bf r}^i_{dd}, \{\hat{{\bf R}}^i\} \right)
~~\rightarrow~~
\Psi_i^L({\bf q}^i,{\bf r}^i_{dd})
\end{equation}

\begin{equation}
F_{3,1}^L\left (\hat{{\bf R}}^f_{3He},{\bf r}^f_4,\{\hat{{\bf R}}^f\} \right )
~~\rightarrow~~
\Psi_f^L({\bf q}^f,{\bf r}^f_4)
\end{equation}

\noindent
The generalized lattice wavefunctions include an explicit dependence on phonon
coordinates, as well as an explicit dependence on a microscopic nuclear degree
of freedom.  In the initial state, this extra degree of freedom is the deuteron-deuteron
separation; in the final state it is the position of the neutron.  Implicit in this
description is an assumption that the lattice center of mass is preserved, which is
an additional assumption here that is consistent with our focus on phonon exchange
(we do not mean to imply here that no coupling is possible with the lattice center of mass
coordinate).
With this notation, we may write

{\small

\begin{equation}
I ~=~
\int  
\left[ \Psi_f^L({\bf q}^f,{\bf r}^f_4) \right]^*
\hat{v}_{fi}({\bf q}^f,{\bf q}^i)
\Psi_i^L({\bf q}^i,{\bf r}^i_{dd})
      \delta^{(3N-3)}( {\bf q}^f - {\bf A} \cdot {\bf q}^i - {\bf b} ) 
d^3{\bf r}^f_4 \; 
d^3 {\bf r}^i_{dd}  
d {\bf q}^f
d {\bf q}^i
\end{equation}

}

\noindent
where the interaction potential is

{\small

$$
\hat{v}_{fi}({\bf q}^f,{\bf q}^i)
~=~
\int
\left[ \phi_{3He}(\xiv^f_{32},\xiv^f_{21})\right]^* \; 
V(\xiv^f_{21}) 
\phi_d(\xiv^i_{21})\, 
\phi_d(\xiv^i_{43}) 
\ \ \ \ \ \ \ \ \ \ \ \ \ \ \ \ \ \ \ \ \ \ \ \ \ \ \ \ \ \ \ \ \ \
\ \ \ \ \ \ \ \ \ \ \ \ \ \ \ \ \ \ \ \ \ \ \ \ \ \ \ \ \ \ \ \ \ \
$$
$$
\delta^{(3)} \left( - \frac{1}{3} \xiv^f_{32} - \frac{2}{3} \xiv^f_{21}
+ \hat{{\bf R}}^f_{3He} - ( - \frac{1}{2} \xiv^i_{21} + \hat{{\bf R}}^i_{4He} -
\frac{1}{2} {\bf r}^i_{dd}) 
\right) 
$$
$$
\delta^{(3)}\left( - \frac{1}{3} \xiv^f_{32} + \frac{1}{3}
\xiv^f_{21} + \hat{{\bf R}}^f_{3He} - (\frac{1}{2} \xiv^i_{21} + \hat{{\bf
  R}}^i_{4He} - \frac{1}{2} {\bf r}^i_{dd}) \right) 
$$
$$  
\delta^{(3)}\left( \frac{2}{3} \xiv^f_{32} + \frac{1}{3} \xiv^f_{21} +
      \hat{{\bf R}}^f_{3He} - (-\frac{1}{2} \xiv^i_{43} + \hat{{\bf
      R}}^i_{4He} + 
      \frac{1}{2} {\bf r}^i_{dd} ) 
      \right) 
$$
\begin{equation}
\delta^{(3)}\left({\bf r}^f_4 - (\frac{1}{2} \xiv^i_{43} + \hat{{\bf
    R}}^i_{4He} + \frac{1}{2} {\bf r}^i_{dd}) \right) \;
d^3\xiv^f_{32} \; 
d^3\xiv^f_{21} \; 
d^3 \xiv^i_{21}\; 
d^3 \xiv^i_{43} \; 
\label{Bigresult1}
\end{equation}

}

\noindent
We may integrate over the final state phonon coordinates to obtain

\begin{equation}
I ~=~
\int  
\left[ \Psi_f^L({\bf q}^f,{\bf r}^f_4) \right]^*
\hat{v}_{fi}({\bf q}^f,{\bf q}^i)
\Psi_i^L({\bf q}^i,{\bf r}^i_{dd})
d^3{\bf r}^f_4 \; 
d^3 {\bf r}^i_{dd}  
d {\bf q}^i
\label{Bigresult2}
\end{equation}

\noindent
in which the final state phonon coordinates are expressed in terms of initial state coordinates
as

\begin{equation}
{\bf q}^f 
~=~ 
{\bf A} \cdot {\bf q}^i + {\bf b}
\end{equation}

\noindent
Equations (\ref{Bigresult1}) and (\ref{Bigresult2}) are the primary results 
that we find in the particular integral involved in the specific
deuteron-deuteron matrix element considered as our example.  Similar results
can be readily obtained for other integrals appearing in this matrix element,
and also for other matrix elements as well.

\newpage
\section{Conclusions}

   The analysis presented in this paper is motivated by claims of excess power production
in experiments involving metal deuterides that have been reported over the years since
the original claim of Fleischmann and Pons in 1989.  
The existence of such an effect implies new physical mechanisms, in particular mechanisms 
which lead to a coupling between condensed matter degrees of freedom and nuclear degrees of freedom.  
There has been confusion and conjecture about how energy or angular momentum exchange 
with the lattice works, and no relevant formulation or calculations have been reported in the
mainstream literature. 
In recent years, our focus has been on a particular scheme involving excitation transfer, idealized
models for which are discussed in detail in a recent manuscript.\cite{PLHandIUC}
The formulation discussed in this paper leads to formulas that can be used for a quantitative
evaluation of this model, and other models that have been proposed which also involve coupling
with phonons.
%
%
%
%
 
The nuclear part of the calculation requires spin and isospin degrees of freedom in addition
to the spatial degrees of freedom.  
We developed nuclear wavefunctions based on Yamanouchi
basis functions for each degree of freedom individually, which allows us to integrate out
the spin and isospin degrees of freedom separately in a convenient way.  
This leads to a straightforward expansion of the interaction matrix element in terms of
a sum of spatial integrals which contain different permutations of particle numbering.

  The lattice part of the problem comes into the calculation initially as spectator degrees
of freedom in the generalized channel separation factor.  
In essence, the microscopic part of the calculation at this level is completely equivalent 
to the vacuum calculation, since the spectator nuclei are not involved locally in the strong force interaction.  
The lattice does come into the problem in a more fundamental way when we replace the nuclear
center of mass degrees of freedom with phonon coordinates.  
Since the reacting nuclei in this formulation are taken to be part of the lattice in both initial 
and final states, the lattice is changed as a result of the reaction.  
We chose a description in which both the initial and final state product nuclei are assumed to 
be included in the associated lattice definitions.
This leads to integrals which contain both initial state and final state lattice wavefunctions,
where the lattice structure is different in the two cases.

%
%

\newpage

\appendix


\section{The nuclear potential}

We make use of the Hamada-Johnston potential as our
model nuclear potential in this paper. 
It is one of the simplest isospin-preserving nuclear
potentials to give reasonable results. 
The Hamada-Johnston potential between nucleon 1 and nucleon 2 can be written as 

\begin{equation}
V_{HJ}(1,2) ~=~ V_C(1,2) ~+~ V_T(1,2) S_{12} ~+~ V_{LS}(1,2) ({\bf
  L}\cdot{\bf S}) ~+~ V_{LL}(1,2) L_{12} 
\end{equation}

\noindent 
where ${\bf L}$ and ${\bf S}$ are the relative angular momentum
and and the total spin of the two nuclei. The individual potentials are
defined as 

\begin{eqnarray*}
V_C(1,2) & = & (\tauv_1 \cdot \tauv_2) \; (\sigmav_1 \cdot \sigmav_2) \; 
y_C(\mu r_{12}) \\
V_T(1,2) & = & (\tauv_1 \cdot \tauv_2) \, y_T(\mu r_{12}) \\
 S_{12} & = & 3\frac{(\sigmav_1 \cdot {\bf r}_{12})
\; (\sigmav_2.{\bf r}_{12})}{r_{12}^2} - \sigmav_1 \cdot
  \sigmav_2 \\
V_{LS}(1,2) & = & y_{LS}(\mu r_{12}) ({\bf L} \cdot {\bf S}) \\
 V_{LL}(1,2) & = & y_{LL}(\mu r_{12})  \\
L_{12} & = &  (\sigmav_1 \cdot \sigmav_2) {\bf L}^2 - 
\frac{1}{2} (\sigmav_1 \cdot {\bf L}) (\sigmav_2 \cdot {\bf L}) -
\frac{1}{2}
(\sigmav_2 \cdot {\bf L}) (\sigmav_1 \cdot {\bf L})
\end{eqnarray*}

The spatial functions are given by 
{\small

\begin{eqnarray*}
y_C^{\alpha}(x) & = & 0.08 \frac{m_\pi c^2}{3} Y(x) \{1 + a_C^\alpha Y(x) + b_C^\alpha Y^2(x)\} \\
y_T^{\alpha}(x) & = & 0.08 \frac{m_\pi c^2}{3} Z(x) \{1 + a_T^\alpha
Y(x) + b_T^\alpha Y^2(x)\} \\
y_{LS}^{\alpha}(x) & = & m_\pi c^2 G_{LS}^\alpha  Y^2(x) \{1 +
b_{LS}^\alpha  Y(x)\} \\
y_{LL}^{\alpha}(x) & = & m_\pi c^2 G_{LL}^\alpha x^{-2} Z(x) \{1 +
a_{LL}^\alpha Y(x) + b_{LL}^\alpha
Y^2(x)\} 
\end{eqnarray*}
}

\noindent 
where $m_\pi$ is the pion mass, ${\alpha}$ stands for odd-singlet($os$) ,
even-singlet($es$), odd-triplet($ot$) or even-triplet($et$).  The functions $Y(x)$ and $Z(x)$ 
have the definitions

$$
Y(x)  ~=~ \frac{e^{-x}}{x}  
\ \ \ \ \ \ \ \ \ \ 
Z(x)  ~=~ \left(1 + \frac{3}{x} + \frac{3}{x^2} \right) Y(x)
$$

\newpage


\section{Symmetric group}
\label{sec:symgroup}

The symmetric group $S(n)$, the group of permutations of $n$ objects,
plays an important role in our calculation. 
It does so, in three important ways
\begin{itemize}
\item Clebsch-Gordan coefficients are used to construct the
  antisymmetric wavefunctions. 
\item Schur-Weyl duality lets us build wavefunctions with a
  well-defined spin (and isospin) and which also transform as 
  Yamanouchi basis vectors under $S(4)$.  
\item Induction coefficients of $S(4)$ are utilized to construct
  Yamanouchi basis vectors for the spatial part.
\end{itemize}

\noindent
In quantum mechanics, we are familiar with the transformation properties of
the spherical harmonics. In the case of $Y_{lm}$'s the $l$ labels the
irreducible representation of $SO(3)$
  and $m$ labels the basis
vector. Similarly for the symmetric group, we get two labels: the Young
diagrams and the Yamanouchi symbols \cite{Hamermesh,Stancu}, where the Young
diagrams label
the irreducible representation of $S(4)$ and the Yamanouchi symbols
label the basis vectors. The index scheme we use
for the Yamanouchi symbols is 
defined in Table~\ref{tbl:YamanouchiShortHand}.   

\begin{table}[h]
\caption{Definition of indices for the Yamanouchi symbols}
\begin{tabular}{|c|c|c|c|} \hline
Yamanouchi Symbol & index & Yamanouchi symbol & index\\ \hline
4321 & 1 & 2121 & 6 \\ 
3211 & 2 & 2111 & 7 \\ 
3121 & 3 & 1211 & 8 \\ 
1321 & 4 & 1121 & 9 \\ 
2211 & 5 & 1111 & 10 \\ \hline
\end{tabular}
\label{tbl:YamanouchiShortHand}
\end{table}

\subsection{Clebsch-Gordan coefficients}
The general formula for constructing antisymmetric wavefunctions is given
by 

{\small

\begin{eqnarray}
\label{eq:CG}
\Psi([f],[f^\prime],[f^{\prime \prime}]) & = & \sum_{Y,\tilde{Y}}
C^{[1^4] ,1}_{[\tilde{f}],\tilde{Y}; [f], Y} R([\tilde{f}],\tilde{Y})
\sum_{Y^\prime,Y^{\prime \prime}}  
C^{[f],Y}_{[f^\prime]Y^\prime;[f^{\prime \prime}]Y^{\prime
      \prime}}\; \; S([f^{\prime}], Y^\prime)\; T([f^{\prime \prime}], Y^{\prime
  \prime})
\end{eqnarray}

}

\noindent
where the $C$'s are the Clebsch-Gordan coefficients of $S(4)$,  $[f]$'s
are the Young diagrams, $Y$'s are the Yamanouchi symbols, $\tilde{f}$ and
$\tilde{Y}$ are the conjugate Young diagrams and Yamanouchi
symbols and $[1^4]$ represents the completely antisymmetric one dimensional
representation with one Yamanouchi basis vector (index 1).  The
Clebsch-Gordan coefficients of $S(4)$ are
available in the literature\cite{Hamermesh,Stancu,Stancu2,Chen}.   

As a specific example, we construct the basis function $\Psi_i =
\Psi_6$.  Using 
our indexing scheme of Table~\ref{tbl:YamanouchiShortHand} above and
Table 4.13 of Chen et al\cite{Chen}, we see that  

\begin{equation}
C^{5}_{10,5} = 1
\ \ \ \ \
C^{6}_{10,6} =  1 
\end{equation}

\noindent
Hence the spin-isospin part of Equation~\ref{eq:CG} yields

\begin{equation}
(st)_5 = s_{10} t_5 
\ \ \ \ \
(st)_6 = s_{10} t_6
\end{equation}

\noindent
where the subscripts in $(st)_5$ or $(st)_6$  again refer to the
Yamanouchi basis for the particular spin-isospin combination. 
In addition, we have

\begin{equation}
C^{1}_{5,6} = \frac{1}{\sqrt{2}} 
\ \ \ \ \ 
C^{1}_{6,5}  =  - \frac{1}{\sqrt{2}}
\end{equation}

\noindent
Therefore, the totally antisymmetric wavefunction can be written as  

\begin{equation}
\Psi_6  
~=~  
\frac{1}{\sqrt{2}} \psi_5 (st)_6 - \frac{1}{\sqrt{2}} \psi_6 (st)_5 
~=~ \frac{1}{\sqrt{2}} \left[\psi_{5}  s_{10}t_6 -  \psi_{6}s_{10} t_5 \right]
\end{equation}


\subsection{Schur-Weyl duality}
The Schur-Weyl duality relates certain representations of $SU(m)$ or
$GL(m)$ to the representations of $S(n)$.  For our
purposes, since we are only interested in the spin-half (or
isospin-half) case, the Schur-Weyl duality is particularly simple. It
states that there are no spin (or isospin) wavefunctions corresponding to
the index 1,2,3 or 4.  It also allows us to construct the rest of
the wavefunctions very simply. In the usual spectroscopic
notation, with the nucleon number as a subscript and $\sigma$ referring
to spin angular momentum.  

\begin{eqnarray}
s_5 & = & | \sigma_1 \sigma_2 (S_{12} = 1), \sigma_3 (S_{123} =
\frac{1}{2}),\sigma_4 \; S = 0 \rangle \nonumber \\
s_6 & = & | \sigma_1 \sigma_2 (S_{12} = 0),\sigma_3 (S_{123} =
\frac{1}{2}), \sigma_4 \; S = 0 \rangle \nonumber \\
s_7 & = & | \sigma_1 \sigma_2 (S_{12} = 1), \sigma_3 (S_{123} =
\frac{3}{2}),\sigma_4 \; S = 1 \rangle \nonumber \\
s_8 & = & | \sigma_1 \sigma_2 (S_{12} = 1), \sigma_3 (S_{123} =
\frac{1}{2}),\sigma_4 \; S = 1 \rangle \nonumber \\
s_9 & = & | \sigma_1 \sigma_2 (S_{12} = 0), \sigma_3 (S_{123} =
\frac{1}{2}), \sigma_4 \; S = 1 \rangle \nonumber \\
s_{10} & = & |  \sigma_1 \sigma_2 (S_{12} = 1),\sigma_3 (S_{123} =
\frac{3}{2}), \sigma_4 \;S = 2 \rangle 
\end{eqnarray}

\noindent
Here on the RHS, we are suppressing the value of $M_S$ because different
$M_S$ 
values transform in the same way under the symmetric group. Exactly the
same results apply to the isospin wavefunctions, since they 
are also angular momentum 1/2 representations of $SU(2)$.

\subsection{Induction Coefficients}

In the previous subsection we explicitly constructed the spin/isospin
Yamanouchi wavefunctions using angular momentum addition. We still have
to construct the spatial Yamanouchi wavefunctions.  One way to do it to
use induction coefficients of the symmetric group, 
which are tabulated and discussed in Chen et al.\cite{Chen}. The
relevant formula for our purposes is 

\begin{equation}
\psi^{n}([f],Y^f) 
~=~ 
\sum_{Y^g, Y^h,\omega} 
~S^{[f],Y^f}_{[g],Y^g; [h],Y^h;\omega} 
~\psi^{n1}([g],Y^g,\omega_1) 
~\psi^{n2}([h],Y^h,\omega_2) 
\end{equation}

\noindent
Here $n = n1 + n2$, $[f]$ and $Y^f$ are the Young diagram and the
Yamanouchi symbol for an $n$-particle spatial wavefunction, $[g]$ and
$Y^g$ are the Young diagram and Yamanouchi symbol for the $n1$-particle
wavefunction and $[h]$ and $Y^h$ are the Young diagram and Yamanouchi
symbol for the $n2$-particle wavefunction and $\omega =
(\omega_1,\omega_2)$ represents a permutation of $\{1,2,\cdots,n\}$. 

As a concrete example let us consider $n = 4$ case with $n1 = n2  =
2$.  Suppose we want to construct a $\psi_6$, starting with two two-body
spatial wavefunctions, $\phi$ and $\phi^\prime$,  which are
completely symmetric.  Hence the 
summation over the Yamanouchi symbols is trivial and we only need to sum
over various $\omega$'s corresponding to different orderings of
$\{1,2,3,4\}$. We can use Table 
4.17 of Chen et al\cite{Chen} to see that (suppressing the trivial Yamanouchi symbols) 

\begin{equation}
S^6_{\{2,3,1,4\}} = - \frac{1}{2} \ \ \ S^6_{\{1,4,2,3\}} = -
\frac{1}{2} \ \ \ S^6_{\{1,3,2,4\}} =  \frac{1}{2} \ \ \
S^6_{\{2,4,1,3\}} =  \frac{1}{2}  
\end{equation}  
 
\noindent
Hence 

\begin{equation}
\psi_6 = \frac{1}{2} \left[ - \phi(2,3) \phi^\prime(1,4) - \phi(1,4)
  \phi^\prime(2,3) + \phi(1,3) \phi^\prime(2,4) + \phi(2,4)
  \phi^\prime(1,3) \right] 
\end{equation}


We can use this approach to develop a complete set of results needed for the case of two
deuterons as used in this paper.  For the initial state, we may write

{\small
\begin{eqnarray}
\psi_{5} & = & \frac{1}{\sqrt{12}} \left[ 2 \psi(12;34) + 2 \psi(34;21)
  - \psi(23;14) -  \psi(14;23) - \psi(13;24) - \psi(24;13) \right]
\nonumber \\
\psi_{6} & = & \frac{1}{2} \left[-\psi(23;14) - \psi(14;23) + \psi(13;24) + \psi(24;13) \right]
\end{eqnarray}
}


\noindent
where $\psi(12;34)$ is any spatial wavefunction which is symmetric under
the $1\leftrightarrow2$ and $3\leftrightarrow4$ exchange.
For the final state, we are using a three-nucleon body and a one nucleon
body. The Yamanouchi wavefunctions are

{\small
\begin{eqnarray}
\psi_{10} & = & 
\frac{1}{2} \left[\psi(123;4) + \psi(124;3) + \psi(134;2) + \psi(234;1) \right] \nonumber \\
\psi_7 & = & \frac{1}{\sqrt{12}} \left[3 \psi(123;4) - \psi(124;3) -
  \psi(134;2) - \psi(234;1) \right] \nonumber \\ 
\psi_8 & = & \frac{1}{\sqrt{6}} \left[2 \psi(124;3) - \psi(134;2) -
  \psi(234;1) \right] \nonumber \\
\psi_9 & = & \frac{1}{\sqrt{2}} \left[ \psi(134;2) - \psi(234;1) \right]
\end{eqnarray}
}

\noindent
where the $\psi(123;4)$ is any wavefunction symmetric under any
permutation of $\{1,2,3\}$. 

 Hence we see that by using induction coefficients, we can take
 spatial wavefunctions which belongs to certain representations
 of subgroups of $S(4)$ and use them to construct Yamanouchi basis
 vectors  e.g.  for the case of two
 deuterons, we started with $\psi(12;34)$ which is symmetric under the
 exchange of $1\leftrightarrow 2$ or $3 \leftrightarrow 4$, and 
 constructed spatial wavefunctions with Yamanouchi basis
 index 5 or 6.

\newpage


\section{Yamanouchi basis states}

Below we explicitly write down the $T = 0$ basis states.  
Please note that the subscripts on the LHS of the equations are
merely ways of enumerating the possible wavefunctions, where as the
subscripts on the RHS are the indices for the Yamanouchi symbols
(see Table~\ref{tbl:YamanouchiShortHand}).  

\begin{samepage}

\noindent
{\bf S = 0}
\begin{eqnarray*}
\Psi_1 & = &  \psi_{10} \frac{1}{\sqrt{2}} \left[  s_5
   t_6 - 
   s_6  t_5 \right] \\ \Psi_2 & = &  \psi_{1}
  \frac{1}{\sqrt{2}} \left[ s_5  t_5 + 
   s_6  t_6 \right] \\
\Psi_3 & = & -\frac{1}{2}  \psi_{5} \left[ s_5 
  t_6  +  s_6
  t_5 \right] -  \frac{1}{2} \psi_{6} \left[ s_5
   t_5 -  s_6  t_6
  \right] 
\end{eqnarray*}

\end{samepage}

\begin{samepage}
\noindent
{\bf S = 1} 
\begin{eqnarray*}
\Psi_4 & = & \frac{1}{\sqrt{6}}  \psi_{7} \left[ s_8
   t_6 -
  s_9 t_5 \right]  + 
\frac{1}{2
  \sqrt{3}} \psi_{8} 
\left[ \sqrt{2}  s_7  t_6 +  s_8 
  t_6 +  s_9  t_5 
  \right] + 
\\ & & 
\frac{1}{2 \sqrt{3}}  \psi_{9} \left[- \sqrt{2}
   s_7 t_5 +  s_8  t_5 -
  s_9  t_6 \right] \\  
\Psi_5 & = & \frac{1}{2 \sqrt{3}}  \psi_{2} \left[ \sqrt{2}
   s_7  t_6 - s_8  t_6 -  s_9  t_5 
  \right] - 
\\ & &  
\frac{1}{2 \sqrt{3}}  \psi_{3} \left[\sqrt{2}  s_7
   t_5 +  s_8 
   t_5 - s_9  t_6 \right] + 
\frac{1}{\sqrt{6}}  \psi_{4} \left[ s_8  t_5 +
  s_9  t_6 
  \right] 
\end{eqnarray*}
\end{samepage}

\begin{samepage}
\noindent
{\bf S = 2} 
\begin{eqnarray*}
\Psi_6 & = & \frac{1}{\sqrt{2}} \left[\psi_{5}  s_{10}
   t_6 -  \psi_{6}  s_{10} t_5 \right] 
\end{eqnarray*}
\end{samepage}

\newpage

\section{The tensor matrix element}
\label{sec:specificmatelt}

We have calculated a complete set of such Yamanouchi matrix elements for
the $T = 0$ case. 
There are too many cases to be presented here. 
We therefore focus on a specific example by calculating the tensor force
matrix element of $\Psi_1(M_s = 0)$ and $\Psi_6(M_s = 1)$.  
The main issue with all these calculations is that the
parameters in most of the nuclear potentials are fit separately to spin
singlet-triplet and even-odd spatial parity (for example, in the Hamada-Johnston
potential the $y$ functions are parameterized by $a$ and $b$ parameters, which 
depend on the spin and parity).  
The most convenient way of dealing with these issues is to use projection operators to separate the
various parity and spin channels.  
However we can, without the use of projection operators,
directly exploit the properties of the interaction and the wavefunctions to
evaluate these matrix elements.

\subsection{Exchange properties of the interaction and wavefunctions} 

First of all, since we are using completely antisymmetric wavefunctions,
the tensor part of the matrix element calculation simplifies to

\begin{eqnarray}
\left \langle \Psi_f \left | 
\sum_{\alpha < \beta} V_T(\alpha,\beta) S_{\alpha \beta} 
\right | \Psi_i \right \rangle
 & = &
\langle \Psi_1 (0)| \sum_{1 \leq i < j \leq 4} V_T(i,j) S_{ij}
  | \Psi_6(1) \rangle \nonumber \\ & = & 6 \langle \Psi_1(0) | V_T(1,2)
  S_{12} | 
  \Psi_6(1) \rangle 
\end{eqnarray}

\noindent
Here the values 0 and 1 in the parentheses represent the $M_S$ values of
the initial and final wavefunctions. So now, as discussed in the main
text, the tensor interaction can be evaluated by

\begin{samepage}

{\small
$$
\left \langle \Psi_f \left | 
\sum_{\alpha < \beta} V_T(\alpha,\beta) S_{\alpha \beta} 
\right | \Psi_i \right \rangle
~=~
3  
\bigg [ 
\langle s_5 t_6 \psi_{10} | V_T(1,2) S_{12} | s_{10}t_6 \psi_5 \rangle
-
\langle s_5 t_6 \psi_{10} | V_T(1,2) S_{12} | s_{10}t_5 \psi_6 \rangle
\ \ \ \ \ \ \ \ \ \ \ \ \ \ \ \ \ \ \ \ \ \ \ \ \ \ \ \ \ \ \ \ 
\ \ \ \ \ \ \ \ \ \ \ \ \ \ \ \ \ \ \ \ \ \ \ \ \ \ \ \ \ \ \ \ 
$$
\vskip -0.150in
\begin{equation}
\ \ \ \ \ \ \ \ \ \ \ \ \ \ \ \ \ \ \ \ \ \ \ \ \ \ \ \ \
-
\langle s_6 t_5 \psi_{10} | V_T(1,2) S_{12} | s_{10}t_6 \psi_5 \rangle
+
\langle s_6 t_5 \psi_{10} | V_T(1,2) S_{12} | s_{10}t_5 \psi_6 \rangle
\bigg ]
\end{equation}
}
\end{samepage}

\noindent
From the explicit form of $V_T(1,2) S_{12}$, we can see that it is
symmetric under the exchange $ 1 \leftrightarrow 2$, individually for spin,
isospin and space parts.  
If we look at the explicit form of the spin wavefunctions in
Appendix~\ref{sec:symgroup} we can see that since particles 1 and 2 couple to form a triplet, $s_5$,
$s_7$, $s_8$ and $s_{10}$ are even under $1 \leftrightarrow 2$. 
Similarly since particles 1 and 2 couple to form a singlet, $s_6$ and $s_9$ are odd under $1
\leftrightarrow 2$.  
The same applies to the isospin wavefunctions. 
Being even or odd under the exchange $1 \leftrightarrow 2$ in this case has nothing to do with angular 
momentum algebra, but is a property of the Yamanouchi symbol. 
Hence the spatial wavefunctions $\psi_5$, $\psi_7$, $\psi_8$ and
$\psi_{10}$ are even under $1 \leftrightarrow 2$
and $\psi_6$ and $\psi_9$ are odd under $1 \leftrightarrow 2$.  As a
concrete example, this can easily be seen by explicitly looking at the
$\psi_6$ from Appendix~\ref{sec:symgroup}

\begin{equation}
\psi_6 = \frac{1}{2} \left[ - \phi(2,3) \phi^\prime(1,4) - \phi(1,4)
  \phi^\prime(2,3) + \phi(1,3) \phi^\prime(2,4) + \phi(2,4)
  \phi^\prime(1,3) \right] 
\end{equation}

\noindent
This $\psi_6$, under the exchange $1 \leftrightarrow 2$, transforms
to 

\begin{equation}
\psi_6 \rightarrow \frac{1}{2} \left[ - \phi(1,3) \phi^\prime(2,4) - \phi(2,4)
  \phi^\prime(1,3) + \phi(2,3) \phi^\prime(1,4) + \phi(1,4)
  \phi^\prime(2,3) \right] = - \psi_6
\end{equation}

\noindent
Because of the symmetry of the interaction under
$1\leftrightarrow2$, individually for space, spin and isospin parts, we only
get non-zero 
couplings when the space, spin and isospin have the same parity (under
$1\leftrightarrow2$) for both
the initial and the final wavefunction. Hence we can see that 

\vskip -0.150in
$$
\langle s_5 t_6 \psi_{10} | V_T(1,2) S_{12} | s_{10}t_5 \psi_6 \rangle = 0 
$$
$$
\langle s_6 t_5 \psi_{10} | V_T(1,2) S_{12} | s_{10}t_6 \psi_5 \rangle = 0 
$$
$$
\langle s_6 t_5 \psi_{10} | V_T(1,2) S_{12} | s_{10}t_5 \psi_6 \rangle = 0
$$ 

\vskip 0.100in
\noindent
The only non-zero matrix element is $\langle s_5 t_6 \psi_{10} |
V_T(1,2) S_{12} | s_{10}t_6 \psi_5 \rangle$.

\subsection{Matrix element calculation}

We have reduced our matrix element calculation to just calculating
$\langle s_5 t_6 \psi_{10} | 
V_T(1,2) S_{12} | s_{10}t_6 \psi_5 \rangle$. We can evaluate the spin
and isospin pieces separately  

\begin{equation} 
\langle t_6 |{\bf \tau}_1.{\bf \tau}_2 |t_6 \rangle ~=~ -3
\end{equation}

\begin{equation}
\langle s_5 (M_S = 0) | S_{12} | s_{10}(M_S = 1) \rangle = - 2 \sqrt{3}
z_{12} \frac{x_{12} + i y_{12}}{r_{12}^2} 
\end{equation}

\noindent
Since the $s_{10}$ and $s_5$ are triplets and $\psi_{10}$ and $\psi_5$
are even, the $y_T(\mu r_{12})$ should be the one parameterized by even
parity and triplet spin. Hence $y_T(\mu r_{12}) = y_T^{et}(\mu r_{12})$ where
$et$ stands for ``even, triplet''.  
 
 We can assemble these results to obtain

{\small

\begin{equation}
\left \langle \Psi_1 (0)\left | \sum_{1 \leq i < j \leq 4} V_T(i,j)
S_{i,j}  \right | \Psi_6(1) \right \rangle 
~=~
18 \sqrt{3} \int \psi_{10}^* \, 
\left [
\frac{ \,(x_{12} + i y_{12})\, z_{12} \, y_T^{et}(\mu r_{12})}{r_{12}^2} 
\right ]
\, \psi_5 ~ d^3 {\bf r}_1 \cdots d^3 {\bf r}_4
\end{equation}

}



\noindent
where $\psi_{10}^* = \psi_{10}^*({\bf r}_1,\cdots,{\bf r}_4)$ and $\psi_5
= \psi_5({\bf r}_1,\cdots,{\bf r}_4)$ are functions of spatial
coordinates alone.

\newpage

\section{Vacuum case antisymmetrization}

In Appendix~\ref{sec:specificmatelt} we saw that it was straightforward to
calculate the matrix element of 
the tensor force in terms of the Yamanouchi basis.  This is in general
true for all nuclear interactions.  Perhaps the simplest
way of carrying out realistic physical calculations is to express 
initial and final state wavefunctions in terms of the Yamanouchi basis.
Then we can use the Yamanouchi basis matrix element results to
calculate the nuclear matrix elements with physical initial and final
states. 
 
For an isospin preserving, $ T = 0$ reaction, in which two deuterons
interact to form a three-nucleon body and a one nucleon
body the relevant states are:

\begin{itemize}
\item $\Psi_{2,2}$:  One initial 2 + 2 state of two deuterons in a
  quintet spin 2  state. We should in general consider the deuterons to be in a
  singlet or triplet as well. However, we are focusing on a single example
  here.  In this enumeration of states, we suppress the various
  $M_S$ values.
\item $\Psi_{3,1}$:  Two final 3 + 1 states (since these can be singlets or
  triplets). These  are a linear superpositions of the $^3H + p$ 
  and $^3He + n$ states.  
\end{itemize}

\subsection{Physical wavefunctions }

   The wavefunctions discussed above are formal Yamanouchi objects which are
very general.  We need to focus on specific wavefunctions for the
different mass 4 ($T=0$) channels in order to proceed.  
For simplicity, we adopt a wavefunction for two deuterons that are
in a quintet (spin 2) state.  In this case, we can use

\begin{equation} 
\Psi_{2,2} ~=~ \mathcal{A}\{\psi(12;34) s_{10} t_6\}
\end{equation}

\noindent
Here, $\mathcal{A}$ is an antisymmetrizer, and $\psi(12;34)$ is the
spatial part of the deuteron wavefunctions. This wavefunction can
be taken to be of the form 

\begin{equation}
\psi(12;34) ~=~ 
\phi_d({\bf r}_2 - {\bf r}_1)\, 
\phi_d({\bf r}_4 - {\bf r}_3) \,
F_{2,2} \left ( \frac{{\bf r}_1 + {\bf r}_2}{2}, 
          \frac{{\bf r}_3 + {\bf r}_4}{2} \right ) 
\end{equation}

  There are two kinds of 3 + 1 wavefunctions, including singlet and
  triplet states.  The singlet $S = 0$   
  states can be written as

\begin{equation}
\Psi_{3,1} = \mathcal{A} \left \{ \psi(123;4) \frac{1}{2} (s_5 t_6 - s_6
t_5) \right \}
\end{equation}


\noindent
The 3+1 triplet $S=1$ states can be written as

\begin{equation}
\Psi_{3,1}^\prime = \mathcal{A} \left \{ \psi(123;4) \frac{1}{2} (s_8 t_6
- s_9 t_5) \right \}
\end{equation}


\noindent
The spatial part of $\Psi_{3,1}$ and $\Psi_{3,1}^\prime$ is of the form  

\begin{equation}
\psi(123;4) ~=~
\phi_{3He}({\bf r}_3 - {\bf r}_2,{\bf r}_2 - {\bf r}_1) \; 
F_{3,1} \left (\frac{{\bf r}_1 + {\bf r}_2 + {\bf r}_3}{3},{\bf r}_4 \right ) 
\end{equation}

\subsection{Physical wavefunctions in terms of Yamanouchi basis states}
\label{inductioncoeff}

All the physical wavefunctions for the initial and final states can be
expressed as linear combinations of the Yamanouchi functions $\Psi_1$,
 $\Psi_4$ and
$\Psi_6$. This is so because of group theoretical considerations and our assumptions that
  the spatial part of the deuteron, triton and helium wavefunctions are
  symmetric under the exchange of any two particles.
The results in particular are 

\begin{equation}
\Psi_{2,2}  ~=~  \Psi_6
\end{equation}

\noindent
with
 
{\small

\begin{eqnarray}
\psi_{5} & = & \frac{1}{\sqrt{12}} \left[ 2 \psi(12;34) + 2 \psi(34;21) -
  \psi(23;14) -  \psi(14;23) - \psi(13;24) - \psi(24;13) \right] \\
\psi_{6} & = & \frac{1}{2} \left[-\psi(23;14) - \psi(14;23) +
  \psi(13;24) + \psi(24;13) \right]
\end{eqnarray}
}

\noindent
In the case of the singlet 3+1 channel, the antisymmetrizer acts to produce a single Yamanouchi
basis state

\begin{equation}
\Psi_{3,1} ~=~  \Psi_1 
\end{equation}

\noindent
where the associated fully symmetric spatial part $\psi_{10}$ can be expressed as

\begin{equation}
\psi_{10} ~=~ 
\frac{1}{2} \left[\psi(123;4) + \psi(124;3) + \psi(134;2)
  + \psi(234;1) \right] 
\end{equation}

\noindent
In the case of the triplet 3+1 channel, the antisymmetrizer acts to
produce a single Yamanouchi basis state

\begin{equation}
\Psi_{3,1}^\prime = \Psi_4 
\end{equation}

\noindent
where the associated mixed symmetry spatial parts $\psi_{7}$, $\psi_8$,
and $\psi_9$ are 

{\small

\begin{eqnarray}
\psi_7 & = & \frac{1}{\sqrt{12}} \left[3 \psi(123;4) - \psi(124;3) -
  \psi(134;2) - \psi(234;1) \right] \\ 
\psi_8 & = & \frac{1}{\sqrt{6}} \left[2 \psi(124;3) - \psi(134;2) -
  \psi(234;1) \right] \\
\psi_9 & = & \frac{1}{\sqrt{2}} \left[ \psi(134;2) - \psi(234;1) \right]
\end{eqnarray}
}

\newpage
\section{Lattice case antisymmetrization }

In this appendix we consider a partial antisymmetrization in the case
of lattice wavefunctions.  We retain the basic vacuum description for
the four nucleons of the vacuum case, and extend the description to
include the other nuclei as spectators.


\subsection{Physical Wavefunctions}

We can implement this program in the case of the initial states to write

\begin{equation} 
\Psi^L_{2,2} ~=~ \mathcal{A}\{\psi(12;34,\{{\bf R}\}) s_{10} t_6\}
\end{equation}

\noindent
Here, $\mathcal{A}$ is the antisymmetrizer acting on particles 1, 2, 3
and 4 only, and $\psi(12;34,\{{\bf R}\})$ is the
spatial part of the deuteron and the lattice wavefunctions. This
wavefunction can be taken to be of the form 

\begin{equation}
\psi(12;34) ~=~ 
\phi_d({\bf r}_2 - {\bf r}_1)\, 
\phi_d({\bf r}_4 - {\bf r}_3) \,
F^L_{2,2} \left ( \frac{{\bf r}_1 + {\bf r}_2}{2}, 
          \frac{{\bf r}_3 + {\bf r}_4}{2},\{{\bf R}\} \right ) 
\end{equation}

  There are two kinds of 3 + 1 wavefunctions, including singlet and
  triplet states.  The singlet $S = 0$   
  states can be written as

\begin{equation}
\Psi^L_{3,1} = \mathcal{A} \left \{ \psi(123;4,\{ {\bf R} \})
\frac{1}{2} (s_5 t_6 - s_6 
t_5) \right \}
\end{equation}


\noindent
The 3+1 triplet $S=1$ states can be written as

\begin{equation}
\Psi_{3,1}^{\prime L} = \mathcal{A} \left \{ \psi(123;4,\{ {\bf R} \}) \frac{1}{2} (s_8 t_6
- s_9 t_5) \right \}
\end{equation}


\noindent
The spatial part of $\Psi^L_{3,1}$ and $\Psi_{3,1}^{\prime L}$ is of the
form

\begin{equation}
\psi(123;4,\{{\bf R}\}) ~=~
\phi_{3He}({\bf r}_3 - {\bf r}_2,{\bf r}_2 - {\bf r}_1) \; 
F^L_{3,1} \left (\frac{{\bf r}_1 + {\bf r}_2 + {\bf r}_3}{3},{\bf r}_4,\{{\bf R}\} \right ) 
\end{equation}

\subsection{Physical wavefunctions in terms of Yamanouchi basis states}

Now since our antisymmetrizer, acts only on the nuclear coordinates, and
not on the $\{{\bf R}\} $, we see that the results of the vacuum case
hold except that the spatial parts of the wavefunctions get augmented.  For
example, we may write

\begin{equation}
\Psi^L_{2,2}  ~=~  \Psi_6
\end{equation}

\noindent 
with

{\small
\begin{eqnarray}
\psi_{5} & = & \frac{1}{\sqrt{12}} \left[ 2 \psi(12;34,\{{\bf R}\} ) +
  2 \psi(34;21,\{{\bf R}\} ) - 
  \psi(23;14,\{{\bf R}\} ) -  \psi(14;23,\{{\bf R}\} ) - 
\right. \\ & & \left.
\psi(13;24,\{{\bf R}\} )   -
  \psi(24;13,\{{\bf R}\} )  \right] \\
\psi_{6} & = & \frac{1}{2} \left[-\psi(23;14,\{{\bf R}\} ) -
  \psi(14;23,\{{\bf R}\} ) + 
  \psi(13;24,\{{\bf R}\} ) + \psi(24;13,\{{\bf R}\} ) \right]
\end{eqnarray}
}

\noindent
In the case of the singlet 3+1 channel, we obtain a similar connection to the
Yamanouchi basis as in the vacuum case

\begin{equation}
\Psi^L_{3,1} ~=~  \Psi_1 
\end{equation}

\noindent 
where the spatial part is now augmented to include the spectator nuclei

{\small
\begin{equation}
\psi_{10} ~=~ 
\frac{1}{2} \left[\psi(123;4,\{{\bf R}\} ) + \psi(124;3,\{{\bf R}\} )
  + \psi(134;2,\{{\bf R}\} ) 
  + \psi(234;1,\{{\bf R}\} ) \right] 
\end{equation}
}

\noindent
The triplet 3+1 channel becomes

\begin{equation}
\Psi_{3,1}^{\prime L} = \Psi_4 
\end{equation}

\noindent 
where 

{\small

\begin{eqnarray}
\psi_7 & = & \frac{1}{\sqrt{12}} \left[3 \psi(123;4,\{{\bf R}\} ) -
  \psi(124;3,\{{\bf R}\} )  -
  \psi(134;2,\{{\bf R}\} ) -  \psi(234;1,\{{\bf R}\} ) \right] \\ 
\psi_8 & = & \frac{1}{\sqrt{6}} \left[2 \psi(124;3,\{{\bf R}\} ) -
  \psi(134;2,\{{\bf R}\}  ) -
  \psi(234;1,\{{\bf R}\} ) \right] \\
\psi_9 & = & \frac{1}{\sqrt{2}} \left[ \psi(134;2,\{{\bf R}\} ) -
  \psi(234;1,\{{\bf R}\} ) \right]
\end{eqnarray}

}

\newpage
\section{Integrals resulting from permutations}

Due to the antisymmetrization, we generate a total of twenty four terms
in the matrix element example in the main body of the paper.  These are 

{\small
\begin{eqnarray}
\label{eq:allterms}
\lefteqn{ \sqrt{12} \; \langle \Psi_{3,1}(M_s = 0) | V(1,2) | \Psi_{2,2}(M_s
  = 1) \rangle ~=~ } \nonumber \\ & &  
\int \psi(123;4)^* v(1,2) \psi(12;34) 
+   
\int \psi(123;4)^* [v(1,2) - \frac{1}{2} v(1,4) - \frac{1}{2}
  v(4,1)] \psi(12;43)  
 \nonumber \\
 & & + 
 \int \psi(123;4)^* [v(1,4) - v(1,2)] \psi(14;23)  + 
 \int \psi(123;4)^* v(4,1) \psi(41;23) + 
 \int \psi(123;4)^* v(1,2) \psi(34;12) \nonumber \\ & & +  
 \int \psi(123;4)^* [v(1,2) - \frac{1}{2} v(1,4) - \frac{1}{2} v(4,1)]
  \psi(43;12) + \int \psi(123;4)^* [ - v(1,2) + v(1,4)] \psi(23;14) 
  \nonumber \\ & & + 
 \int \psi(123;4)^* v(4,1) \psi(23;41) +
 \int \psi(123;4)^* [- v(1,2)] \psi(24;13)  \nonumber \\ & & + 
 \int \psi(123;4)^* [- \frac{1}{2} v(1,4) - \frac{1}{2}  v(4,1)]
  \psi(42;13) +  
 \int \psi(123;4)^* [- v(1,2)] \psi(13;24) \nonumber \\ & & + 
 \int \psi(123;4)^* [- \frac{1}{2} v(1,4) - \frac{1}{2} v(4,1)] \psi(13;42) 
\end{eqnarray}
}

\noindent
where $V(1,2)$ is some general nuclear potential which contains spin and
isospin dependent terms, and where $v(i,j)$ is the spatial interaction between
particles $i$ and $j$ which is obtained by carrying out the spin and isospin
algebra. In the special case that $V(1,2) = V_{HJ}(1,2)$, then $v(i,j) =
V^{10,5}_R(i,j)$.  

Some of these terms are amenable to a simple
interpretation in terms of stripping reactions. Others can be considered
as higher-order exchange terms.  However, in nuclear reactions, it is
not necessary that higher
order exchange terms are necessarily less important than simple direct
or exchange interactions.  Hence, in any realistic computation we will
need to sum over all of these terms.



\newpage

\section*{Acknowledgments}
Support for I. Chaudhary was provided by the Kimmel Fund, the Bose Foundation,
the Griswold Gift Program and by a Darpa subcontract.

\end{document}